\documentclass[structabstract]{aa}

\usepackage{amssymb}
\usepackage[usenames, dvipsnames]{color}
\usepackage{graphicx}
\usepackage{natbib}
\usepackage[rightcaption]{sidecap}
\usepackage{txfonts}
\usepackage{units}
\usepackage{url}
\usepackage{pifont}
\usepackage[modulo, switch]{lineno}
\usepackage{hyperref}
\hypersetup{
    final=true,
    pageanchor=true,
    colorlinks=true,
    breaklinks=true,
    linkcolor=blue,
    citecolor=blue,
    urlcolor=blue,
    pdfpagemode=UseNone,
    pdftitle={High-resolution imaging spectroscopy of micro-pores 
    in a small emerging-flux region.},
    pdfauthor={S.J.\ Gonzalez Manrique and C. Denker},
    pdfsubject={Solar Physics},
    pdfkeywords={Sun: chromosphere, Sun: activity, Sun: filaments,
    Techniques: high angular resolution, Methods: observational, 
    Instrumentation: interferometers}}

\definecolor{blue}{rgb}{0.00, 0.00, 1.00}
\definecolor{red}{rgb}{0.86, 0.08, 0.24}
\definecolor{orange}{rgb}{1.00, 0.55, 0.00}
\definecolor{darkblue}{rgb}{0.00, 0.00, 0.55}
\definecolor{green}{rgb}{0.00, 0.39, 0.00}
\definecolor{pink}{rgb}{1.000000,0.078431,0.576471}

\newcommand\tsp{\mbox{$\;\!$}}

\newcommand\phn{\phantom{0}}
\newcommand\phnn{\phantom{00}}

\newcommand\phm{\phantom{$-$}}

\begin{document}


\title{High-resolution imaging spectroscopy of two micro-pores\\
    and an arch filament system in a small emerging-flux region}
\titlerunning{High-resolution imaging spectroscopy of two micro-pores}

\author{S.J.\ Gonz{\'a}lez Manrique\inst{1,2}, N.\ Bello Gonz{\'a}lez\inst{3}
    \and C.\ Denker\inst{1}}

\institute{$^1$ Leibniz-Institut f{\"u}r Astrophysik Potsdam (AIP),
    An der Sternwarte 16, 
    14482 Potsdam, Germany\\
    $^2$ Universit{\"a}t Potsdam, Institut f{\"u}r Physik and
    Astronomie, Karl-Liebknecht-Stra{\ss}e 24/25,
    14476 Potsdam-Golm, Germany\\
    $^3$ Kiepenheuer-Institut f{\"u}r Sonnenphysik, 
    Sch{\"o}neckstra{\ss}e 6, 79104 Freiburg, Germany\\
    \email{smanrique@aip.de, nbello@leibniz-kis.de, cdenker@aip.de}}

\date{Received 2 December 2015; Accepted 22 December 2016}

\abstract
{Emerging flux regions mark the first stage in the accumulation of magnetic flux 
eventually leading to pores, sunspots, and (complex) active regions. These flux 
regions are highly dynamic, show a variety of fine structure, and in many 
cases live only for a short time (less than a day) before dissolving quickly into the 
ubiquitous quiet-Sun magnetic field.}
{The purpose of this investigation is to characterize the temporal evolution of 
a minute emerging flux region, the associated photospheric and chromospheric flow 
fields, and the properties of the accompanying arch filament system. We aim to explore 
flux emergence and decay processes and investigate if they scale with structure size and magnetic
flux contents.}
{This study is based on imaging spectroscopy with the G{\"o}ttingen 
Fabry-P{\'e}rot Interferometer at the Vacuum Tower Telescope, Observatorio del 
Teide, Tenerife, Spain on 2008 August~7. Photospheric horizontal proper motions 
were measured with Local Correlation Tracking using broadband images restored 
with multi-object multi-frame blind deconvolution. Cloud model (CM) inversions 
of line scans in the strong chromospheric absorption H$\alpha$ 
$\lambda656.28$~nm line yielded CM parameters (Doppler velocity, Doppler width, 
optical thickness, and source function), which describe the cool plasma 
contained in the arch filament system.}
{The high-resolution observations cover the decay and convergence of two 
micro-pores with diameters of less than one arcsecond and provide decay 
rates for intensity and area. The photospheric horizontal flow speed is 
suppressed near the two micro-pores indicating that the magnetic field is 
already sufficiently strong to affect the convective energy transport. The 
micro-pores are accompanied by a small arch filament system as seen in 
H$\alpha$, where small-scale loops connect two regions with H$\alpha$ line-core 
brightenings containing an emerging flux region with opposite polarities. The 
Doppler width, optical thickness, and source function reach the largest values 
near the H$\alpha$ line-core brightenings. The chromospheric velocity of the 
cloud material is predominantly directed downwards near the footpoints of the 
loops with velocities of up to 12~km~s$^{-1}$, whereas loop tops show upward 
motions of about 3~km~s$^{-1}$. Some of the loops exhibit signs of twisting 
motions along the loop axis.}
{Micro-pores are the smallest magnetic field concentrations leaving a photometric 
signature in the photosphere. In the observed case, they are accompanied by a 
miniature arch filament system indicative of newly emerging flux in the form of 
$\Omega$-loops. Flux emergence and decay take place on a time-scale of about 
two days, whereas the photometric decay of the micro-pores is much more rapid 
(a few hours), which is consistent with the incipient submergence of 
$\Omega$-loops. Considering lifetime and evolution timescales, impact on the 
surrounding photospheric proper motions, and flow speed of the chromospheric 
plasma at the loop tops and footpoints, the results are representative for the 
smallest emerging flux regions still recognizable as such.}
 

\maketitle


\section{Introduction}\label{SEC1}
Solar activity is intimately linked to the presence of magnetic fields on the 
solar surface. Small-scale magnetic fields emerge within the 
interior of supergranular cells, and they are transported by a radial flow 
pattern to the boundaries of the cells \citep{Martin1988}. The stronger magnetic 
fields appear in a multi-stage process eventually leading to pores, sunspots, 
and (complex) active regions \citep{Zwaan1985, Rezaei2012}.
The first indications of newly emerging flux are ephemeral 
regions with dimensions of about 30~Mm and a total magnetic flux of up to 
$10^{20}$~Mx. These small-scale, bipolar magnetic flux concentrations spring up 
everywhere on the solar surface. They continuously replenish the quiet-Sun magnetic 
field, completely renewingit about every 14~hours \citep{Hagenaar2001, Guglielmino2012}. Emerging 
$\Omega$-loops at granular scale were reported by \citet{MartinezGonzalez2010}. 
When such loops rise trough the photosphere their mean velocity can reach up to 
3~km~s$^{-1}$. Once the loops reach the upper photosphere, 
their magnetic field is almost vertical at the footpoints. In 
the chromosphere, the mean upflow velocity climbs to 12~km~s$^{-1}$, and the 
energy contained in the loops system is at least $1.4 \times 10^{6}$\,--\,$2.2 
\times 10^{7}$~erg~cm$^{-2}$~s$^{-1}$ in the lower chromosphere
\citep{MartinezGonzalez2010}. The emerging fields are transported by 
supergranular flows migrating toward the network \citep{OrozcoSuarez2012}.

\begin{figure}[t]
\includegraphics[width=\columnwidth]{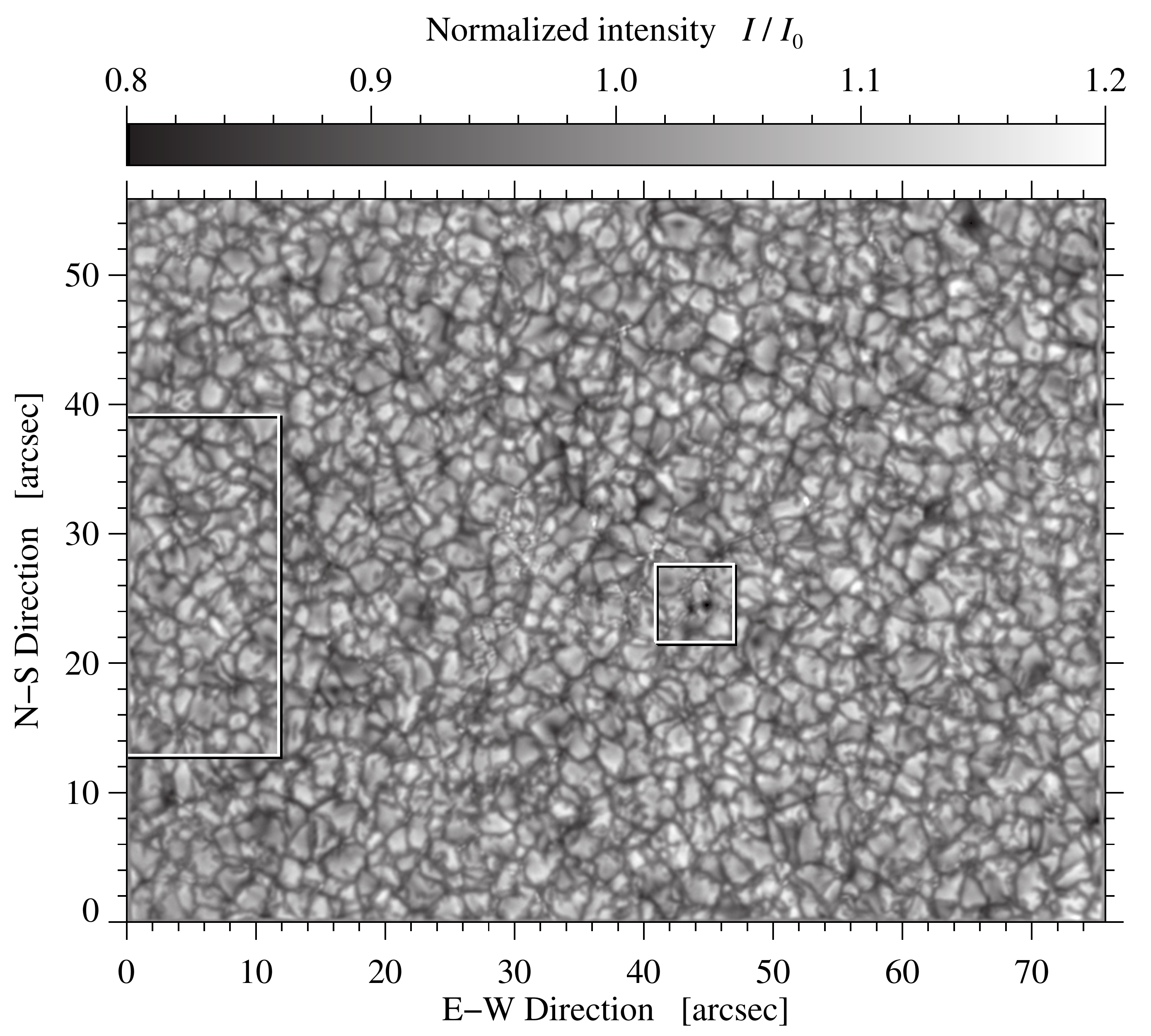}
\includegraphics[width=\columnwidth]{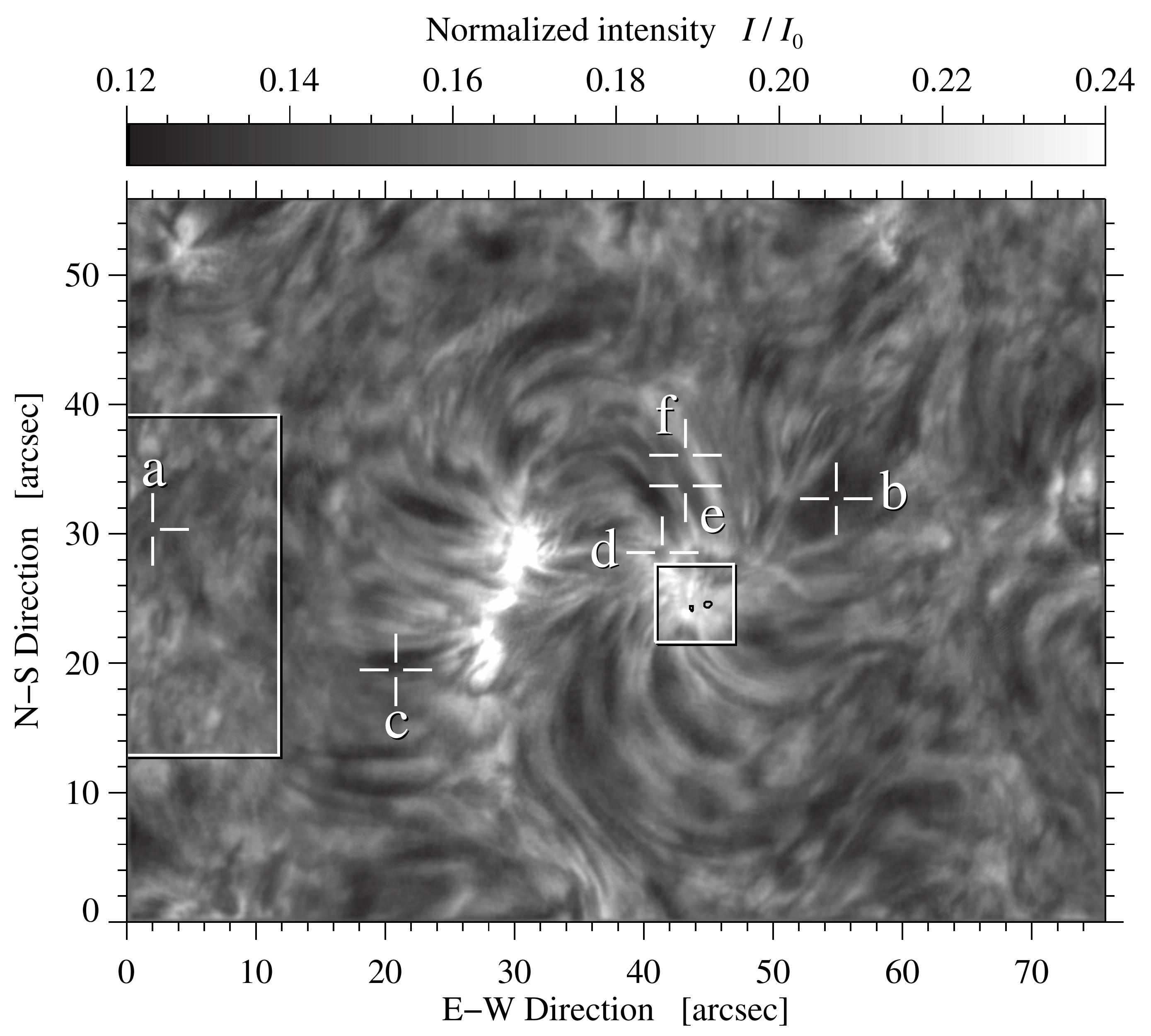}
\caption{MOMFBD-restored broadband image (\textit{top}) at 
    about $\lambda$600~nm of micro-pores (white square in the center in both 
    images) in an EFR observed with the GFPI at 08:07~UT on 2008 August~7. 
    H$\alpha$ line-core intensity image corresponding to the broadband image 
    (\textit{bottom}). The white rectangle on the left outlines the region used 
    for computing the average H$\alpha$ quiet-Sun spectral profile. Crosses 
    `\ding{59}' and alphabetic labels mark the locations of six contrast 
    profiles plotted in Fig.~\ref{FIG_PLOT_CONTRAST_FIT}. The black contours 
    indicate the position of the micro-pores.}
\label{FIG01}
\end{figure}

The initial expansion rate of the bipoles is about 2~km~s$^{-1}$ 
\citep{Harvey1973}. Once the magnetic field becomes sufficiently strong (i.e., 
the filling factor becomes sufficiently high), `micro-pores' appear as a 
photospheric signatures of flux emergence \citep{Scharmer2002, 
RouppevanderVoort2005}. They change their appearance between 
elongated features, which are darker in the center and have their maximum 
brightness at the edges (ribbons), and more circular magnetic structures 
(flowers), adapting the nomenclature of \citet{RouppevanderVoort2005}. 
The definitions were made based on G-band data. Micro-pores 
are at the very low end of the statistical size distribution of pores 
\citep{Verma2014}. Once pores are present, sunspots are eventually formed by 
coalescence of newly emerged flux \citep{Schlichenmaier2010b, Rezaei2012}.

The dynamics of emerging flux regions (EFRs) are characterized by a transient 
state \citep{Zwaan1985} of magnetic field lines rising within granular 
convection. On timescales of about 10~minutes (photospheric crossing time), 
emerging field lines form a pattern of aligned dark intergranular lanes 
\citep{Strous1996}. When reaching chromospheric heights, the magnetic loops 
become visible in H$\alpha$, and material gradually drains from the loops. This 
and convective collapse \citep{Cheung2008} at the footpoints of the loops lead 
to strong downflows.

Arch filament systems (AFS) connecting the two opposite magnetic polarities of 
newly emerging flux are prominently visible in line-core filtergrams of the 
strong chromospheric absorption line H$\alpha$ and also, but less pronounced, in 
the Ca\,\textsc{ii}\,H\,\&\,K lines \citep{Bruzek1969}. Downflows in the range 
of 30\,--\,50~km~s$^{-1}$ occur near both footpoints of dark filaments, whereas 
loop tops rise with about 1.5\,--\,20~km~s$^{-1}$ \citep{Bruzek1969, Zwaan1985, 
Chou1988, Lites1998}. The arched filaments are typically confined below 10~Mm. 
\citet{Bruzek1969} relates the length of the filaments (20\,--\,30~Mm) to the 
size of supergranular network cells. The height of the arches is typically 
5\,--\,15~Mm, and the width of individual filaments is just a few megameters 
with a lifetime of about 30~minutes \citep{Bruzek1967}. However, individual 
loops of the AFS can reach heights of up to 25~Mm and a mean length of about 
20\,--\,40~Mm \citep{Tsiropoula1992}. In general, the appearance of an AFS 
remains the same for several hours, whereas significant changes occur only along 
with the growth of the sunspot group, that is after about three days the AFS 
vanishes.

\begin{figure}[t]
\includegraphics[width=\columnwidth]{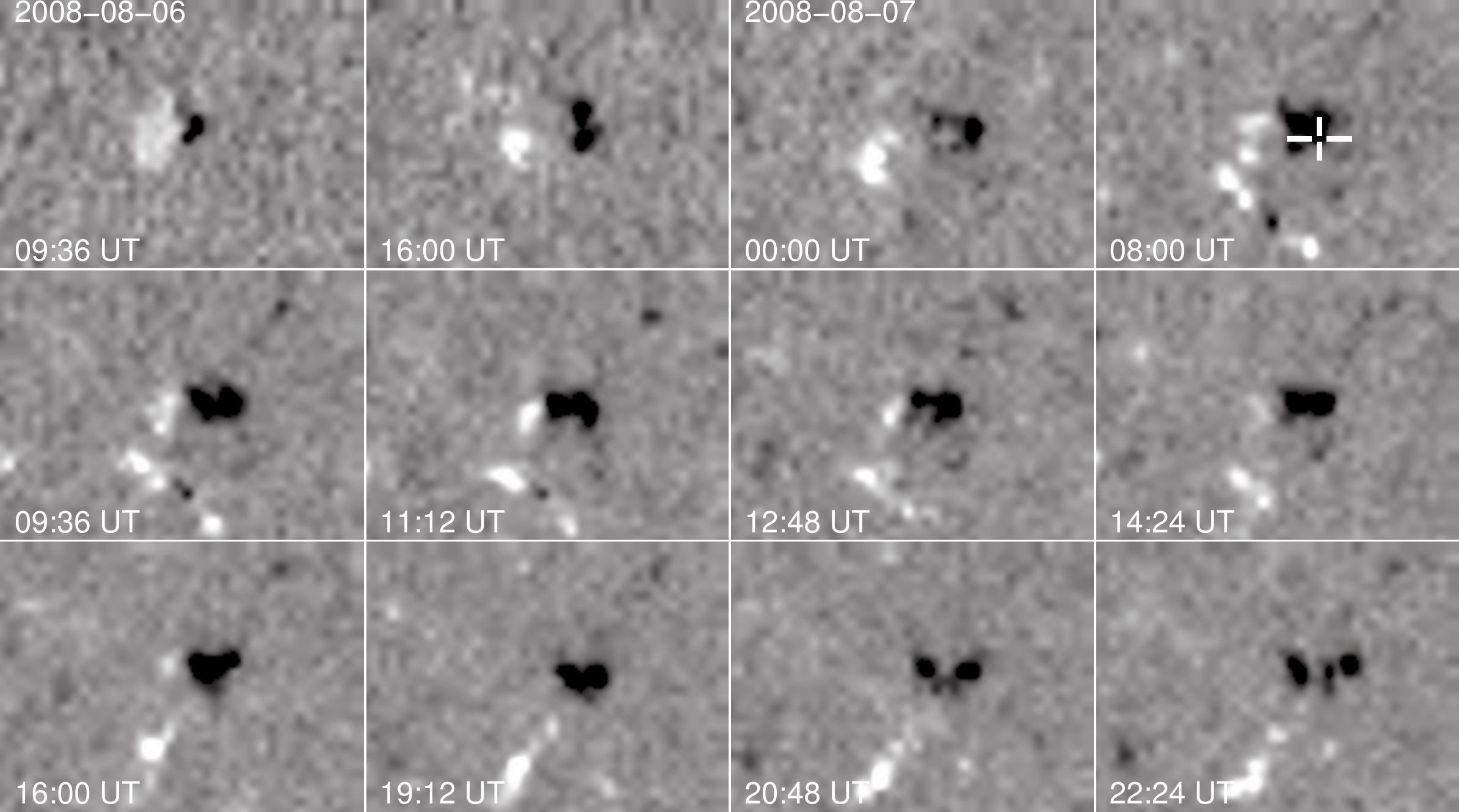}
\caption{SoHO/MDI time series of a small EFR on 2008 August~6 and~7. Each panel 
    has the same size as the high-resolution image in Fig.~\ref{FIG01}. The 
    magnetograms were scaled between $\pm$100~G and resampled to a four times 
    finer grid for better display. The  cross \ding{59}' 
    indicate the position of the micro-pores in Fig.~\ref{FIG01} using the 
    magnetogram closest in time.}
\label{BIPOLE}
\end{figure}

AFS are mainly observed in H$\alpha$, and one method to 
analyze H$\alpha$ spectra is cloud modeling. The cloud model (CM) assumes that 
absorbing chromospheric plasma is suspended by the magnetic field above the 
photosphere \citep{Beckers1964}. A broad spectrum of solar fine structure was 
investigated using CM inversions, for example, H$\alpha$ upflow events \citep{Lee2000}, 
dark mottles in the chromospheric network \citep{Lee2000, Al2004, Contarino2009, 
Bostanci2011}, superpenumbral fibrils \citep{Alissandrakis1990}, AFSs 
\citep{Alissandrakis1990, Contarino2009}, and filaments \citep{Schmieder1991}. 
Such inversions are attractive mainly because of the inherent 
simplicity of the model and the scarcity of other well established inversion 
schemes for strong chromospheric absorption lines such as H$\alpha$ 
\citep[cf.,][]{MolownyHoras1999, Tziotziou2001}. However, refinements of CM 
inversions have been implemented including the differential cloud model 
\citep{Mein1988}, the addition of a variable source function and velocity 
gradients \citep{Mein1996, Heinzel1999}, the embedded cloud model 
\citep{Chae2014}, and the two-cloud model \citep{Hong2014}. \citet{Bostanci2010} 
examined the impact of observed and theoretical quiet-Sun background profiles on 
CM inversions and concluded that synthetic H$\alpha$ profiles based on non-local
thermodynamic equilibrium (NLTE) calculations perform better than profiles derived 
directly from the data.

Physical processes including radiation, convection, conduction, and magnetic 
field generation and decay play an important and prominent role in the solar 
atmosphere. To fully appreciate the dynamics of solar fine structure, 
simultaneous multi-wavelengths observations at different layers of the Sun are 
crucial. \citet{WedemeyerBoehm2009} critically reviewed the links between the 
photosphere, chromosphere, and corona of the Sun. They affirm that these 
atmospheric layers are coupled by the magnetic field at many different spatial 
scales. Thus loops, as the ones found in AFS, reach from the photosphere to the 
chromosphere and corona, where they encounter physical conditions differing by 
orders of magnitude in temperature and gas density. The 
present study strives to contribute to the quantitative description of EFRs and 
AFSs. Some of the key questions related to this study are the following: Do 
morphological evolution and statistical properties of micro-pores imply a 
specific flux emergence scenario or flux dispersion mechanism 
(Sect.~\ref{SEC3.1})? Which physical quantities are most suitable to restrain 
theoretical models? Can the results of CM inversions help to identify distinct 
chromospheric features (cluster analysis in Sect.~\ref{SEC5})?

In Sect.~\ref{SEC2}, we briefly introduce the observations, image processing, 
and image restoration techniques. The data analysis is 
described in Sect.~\ref{SEC5}. Sect.~\ref{SEC3} puts forward the results, where 
we first examine the two micro-pores, then determine the photospheric horizontal 
proper motions associated with the EFR, and finally scrutinize the chromospheric 
response of emerging flux, which includes Doppler velocities, CM inversions, and 
parameters characterizing the AFS. We conclude our study by comparing our 
results with the most recent findings and synoptic literature concerning EFRs 
and AFSs (Sect.~\ref{SEC4}). Initial results were presented in 
\citet{Denker2009}.

%
%

\section{Observations}\label{SEC2}

A small EFR at heliographic coordinates E28.5$^\circ$ and S7.91$^\circ$ ($\mu = 
\cos\theta = 0.86$) was observed on 2008 August~7 with the Vacuum Tower 
Telescope \citep[VTT,][]{vonderLuehe1998} at Observatorio del Teide, Tenerife, 
Spain. Imaging spectroscopy in the strong chromospheric absorption line 
H$\alpha$ $\lambda656.28$~nm was carried out with the G{\"o}ttingen 
Fabry-P{\'e}rot Interferometer \citep[GFPI,][]{Puschmann2006, 
BelloGonzalez2008}. The field-of-view (FOV) of the instrument is $77.1\arcsec 
\times 58.2\arcsec$ ($688 \times 520$ pixels after 2$\times$2-pixel binning) 
with an image scale of 0.112\arcsec\ pixel$^{-1}$. All data were taken under 
good seeing conditions (Fried-parameter $r_{0} > 15$~cm for most of the time) 
with real-time image correction provided by the Kiepenheuer Adaptive Optics 
System \citep[KAOS,][]{vonderLuehe2003,Berkefeld2010}.

Four short time series of simultaneous broad- and narrow-band images were 
acquired with the GFPI during the time period from 07:54\,--\,08:37~UT (see Fig.~\ref{FIG01}). The 
observations were paused twice for about six minutes to take flat-field images 
at solar disk center and continuum images with an artificial light source. The 
AO system experienced a few interruptions while tracking on solar granulation. 
From the 54 time series, 17 data sets (08:07\,--\,08:16~UT) have been selected 
for a more detailed analysis because of very good seeing conditions and 
continuous AO correction. Post-processing with Multi-Object Multi-Frame Blind 
Deconvolution \citep[MOMFBD,][]{Loefdahl2002, vanNoort2005} as described in 
Sect.~4.4 of \citet{delaCruzRodriguez2015} further improves the spatial 
resolution and contrast of the imaging spectroscopic data.

Each spectral scan of narrow-band images covers 61 equidistant 
\mbox{($\delta\lambda = 3.13$~pm)} positions in the strong chromospheric 
absorption line H$\alpha$ $\lambda$656.28~nm, i.e., a spectral window of 
$\Delta\lambda = 0.189$~nm centered on the line core. Eight images were taken at 
each position to increase the signal-to-noise ratio in preparation for MOMFBD. 
With an exposure time of 15~ms, the cadence of a spectral scan with $61 \times 8 
= 488$ single exposures is $\Delta t = 34$~s. Thus, the total duration of the 
selected time series is $\Delta T \approx 10$~min. In some instances, scans of 
the other time series are used, for example, when determining the lifetimes of 
micro-pores. 

Finally, we matched our high-resolution observations to 96-minute cadence, 
full-disk magnetograms of the Solar and Heliospheric Observatory/Michelson 
Doppler Imager \citep[SoHO/MDI][]{Scherrer1995}. The magnetograms have $1024 
\times 1024$ pixels and an image scale of 1.986\arcsec\ pixel$^{-1}$. Thus, the 
image scales of the GFPI and MDI data roughly differ by a factor of twenty, 
that is, a magnetogram section of just $39 \times 29$ pixels corresponds to the 
GFPI FOV. The magnetograms selected for the time series 
depicted in Fig.~\ref{BIPOLE} were chosen taking into account some data gaps and 
the noise being present in the magnetograms. The cadence of the first row is 
about four hours and about 96~minutes in the other two rows.

%
%

\section{Data analysis}\label{SEC5}


\subsection{Local correlation tracking}\label{SEC5.1}

Horizontal proper motions are derived from the time series of 17 broadband 
images using local correlation tracking \citep[LCT,][]{November1988} as 
described in \citet{Verma2011}. However, the data have not been corrected for 
geometrical foreshortening because of the small FOV and its proximity to disk 
center. The input parameters selected for LCT are a cadence of $\Delta t = 
34$~s, an averaging time of $\Delta T \approx 10$~min, and a Gaussian sampling 
window with a $\mathrm{FWHM} = 600$~km (the Gaussian used as a high-pass filter 
has a full-width-at-half-maximum $\mathrm{FWHM} = 1200$~km). The image scale of 
the GFPI broadband and Hinode/BFI G-band \citep{Kosugi2007, Tsuneta2008} images are 
almost the same, thus facilitating a straightforward comparison of flow fields derived from these 
two instruments.


\subsection{Cloud model inversions}\label{SEC5.2}

Some characteristics of strong chromospheric absorption lines are easier to 
discover in intensity contrast profiles, which are given as
\begin{equation}
C(\lambda) = \frac{I(\lambda) - I_0(\lambda)}{I_0(\lambda)}\,,
\label{eq1}
\end{equation}
where $I(\lambda)$ and $I_0(\lambda)$ denote the observed and the average 
quiet-Sun spectral profiles, respectively. \citet{Beckers1964} assumes in his 
model a cloud of absorbing material suspended by the magnetic field above the 
photosphere, and he provides a relationship between the intensity contrast 
profile $C(\lambda)$ and four free fit parameters describing the cloud material, 
that is, the central wavelength of the absorption profile $\lambda_c$, the Doppler 
width of the absorption profile $\Delta\lambda_D$, the optical thickness 
$\tau_0$ of the cloud at the central wavelength, and the source function $S$:
\begin{eqnarray}
C(\lambda)    & = & \left[ \frac{S}{I_0(\lambda)} - 1 \right]
                    \Big( 1 -\exp\big[-\tau(\lambda)\big] \Big)
                    \quad {\rm with} \label{EQN_CM} \\
\tau(\lambda) & = & \tau_0 \exp \left[ - \left(
                    \frac{\lambda-\lambda_c}{\Delta\lambda_D} \right)^2 \right]
\end{eqnarray}
The line-of-sight (LOS) velocity $v_\mathrm{LOS}$ of the cloud can be derived, once $\lambda_c$ 
is known, according to
\begin{equation}
v_\mathrm{LOS} = c\, \frac{\lambda_c - \lambda_0}{\lambda_0}\,,
\end{equation}
where $\lambda_0$ is the central wavelength of the strong chromospheric 
absorption line averaged over a quiet-Sun area and $c$ the speed of light.

The quiet-Sun profile $I_0(\lambda)$ is computed within an about 10\arcsec-wide 
region to the East of the EFR (see white rectangle in Fig.~\ref{FIG01}). 
The selection of $I_0(\lambda)$ has a strong impact on the 
results of the CM analysis \citep{Tziotziou2003}, for example, lateral radiative 
exchange affects the conditions of the cloud material. The selected quiet-Sun 
region is at the periphery of AFS, it does not show any features with strong 
absorption in H$\alpha$, and chromospheric velocities are low. Therefore, this 
quiet-Sun region is the best choice within the available FOV. We carried out 
some tests using different samples within this region to create different 
quiet-Sun profiles. The results from CM inversions were basically identical, 
which is not the case for other `quiet' settings within the FOV, which contained 
either absorption features or exhibited high chromospheric velocities. Line 
shifts are corrected by linear interpolation before averaging the profiles. 
Assuming that up- and downflows in the quiet Sun are balanced and that the 
convective blueshift of spectral lines is negligible in the chromosphere, the 
mean velocity is set to zero and used as a reference for the entire FOV. In this 
study, the quiet-Sun profile $I_0(\lambda)$ is computed from the data itself. 
However, \citet{Bostanci2011} demonstrated that theoretical profiles from NLTE 
calculations may compare favorably to profiles derived directly from 
observations. 

\begin{figure}[t]
\includegraphics[width=\columnwidth]{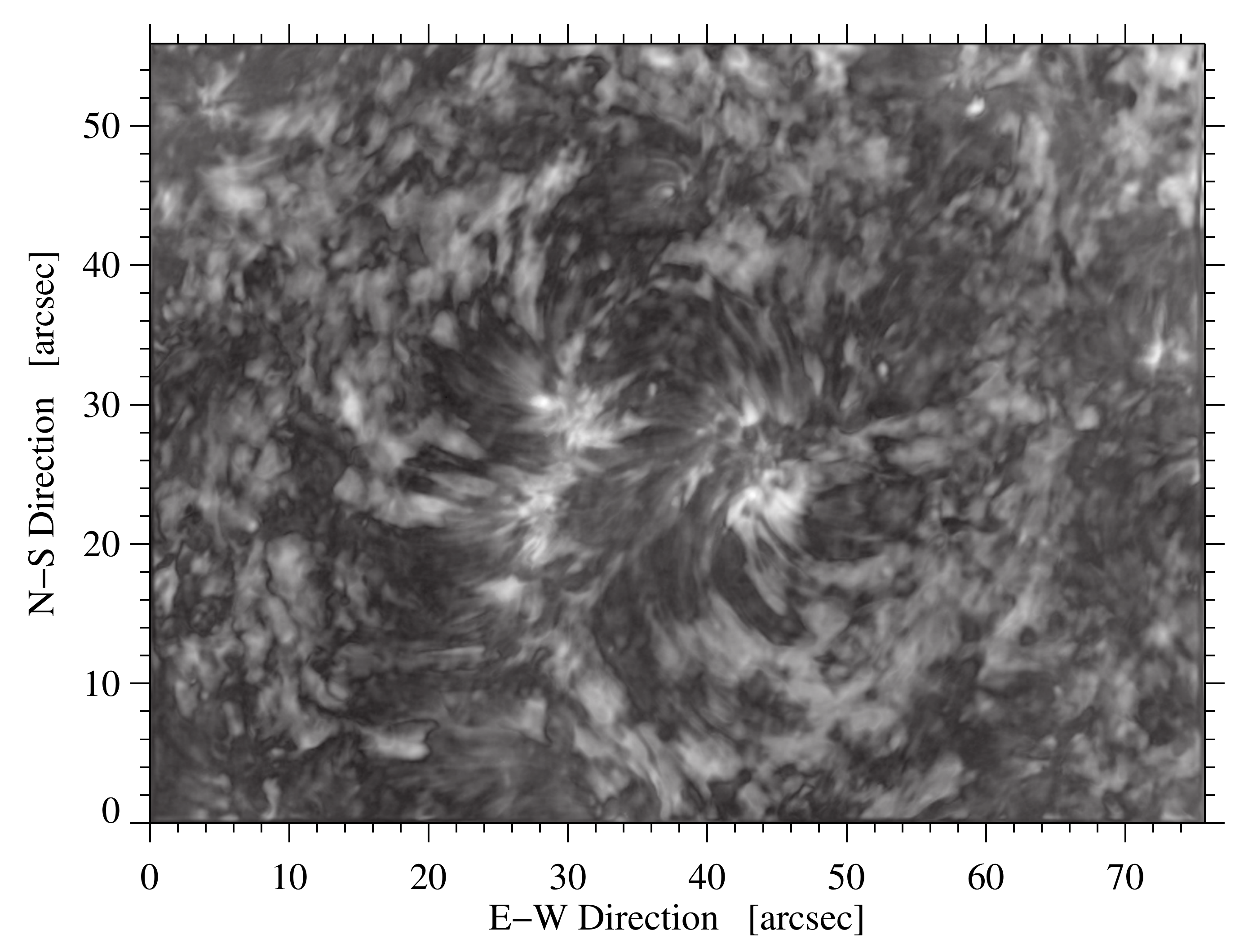}
\caption{Map of the $\chi^2$ goodness-of-fit statistics, where bright regions
    indicate significant differences between the observed and fitted profiles,
    in particular near the footpoints of the AFS.}
\label{CHI2}
\end{figure}

\begin{figure}[t]
\includegraphics[width=\columnwidth]{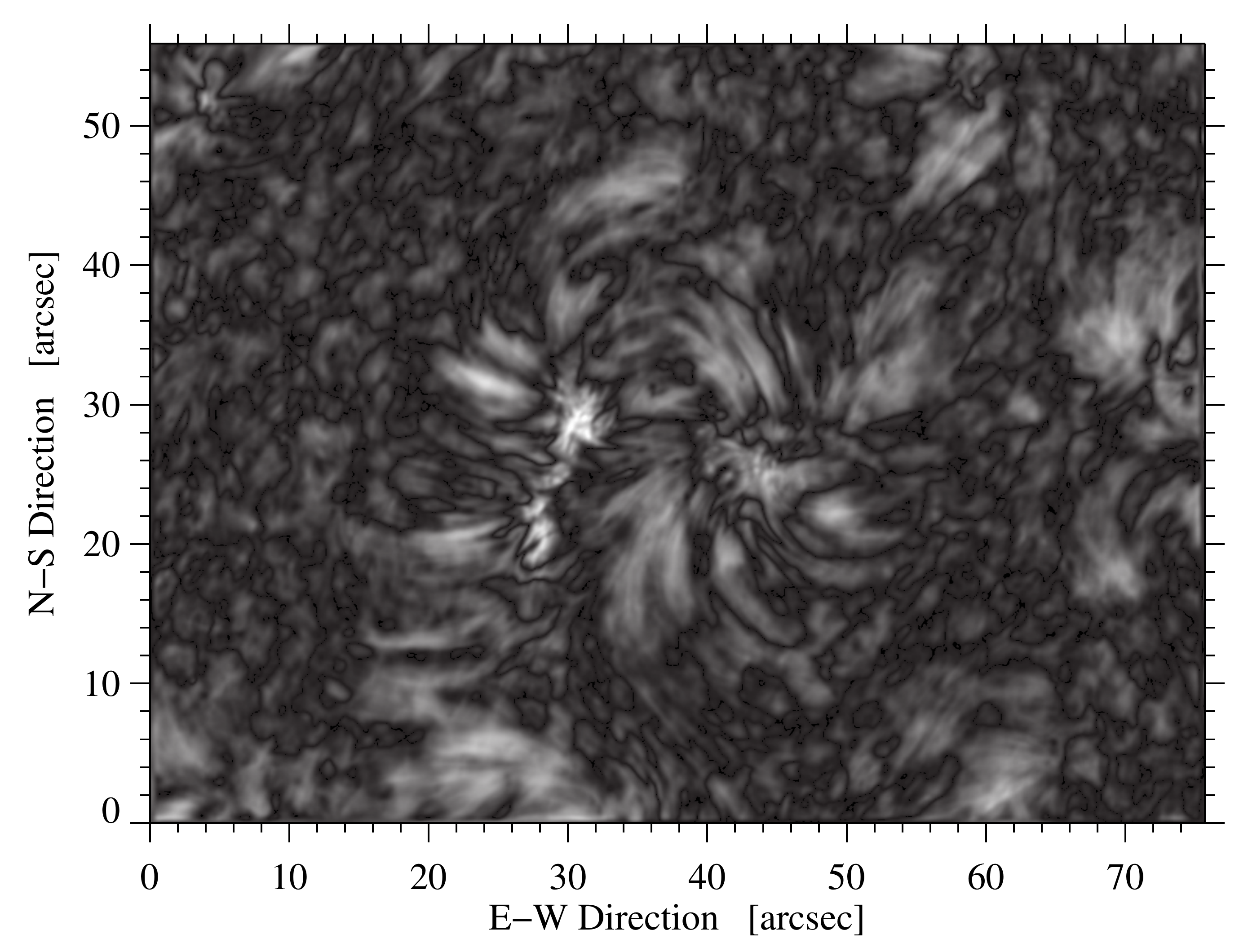}
\caption{Map of the absolute contrast for each spectral profile. The highest
    (positive) contrasts correspond to the bright regions in the H$\alpha$ 
    line-core intensity image shown in Fig.~\ref{FIG01}.}
\label{ABSCON}
\end{figure}

Two steps are required in the CM inversions to efficiently analyze the millions 
of spectral profiles obtained with imaging spectroscopy: (1) Contrast profiles 
are computed for 50\,000 random CM input parameters, which are restricted to the 
intervals $\tau_0 \in [0, 3]$, $v_\mathrm{LOS} \in [-90, +90]$~km~s$^{-1}$, 
$\Delta \lambda_D \in [0, 70]$~pm, and $S \in [0, 0.4]$, where the source 
function $S$ is given in quiet-Sun intensity units. In addition, the 
histograms of the parameters have to closely match those of 
the observations. This requires some a priori knowledge acquired by dropping the 
last assumption for a representative sample of contrast profiles. Each observed 
profile is then compared against the 50\,000 templates, and the CM parameters of 
the closest match (smallest $\chi^2$-values) are saved. (2) The saved CM 
parameters are the initial estimates to perform a Levenberg-Markwardt 
least-squares minimization \citep{More1977, More1993} using the MPFIT software 
package \citep{Markwardt2009}. The quiet-Sun spectral profile is handed to the 
fitting routine as private data avoiding common block variables, and MPFIT's 
diagnostic capabilities provide the means to easily discriminate between 
successful fits and situations, where the iterative algorithm does not converge.

Mediocre fits result in high $\chi^2$-values (Fig.~\ref{CHI2}), which are 
cospatial with the H$\alpha$ line-core brightenings at the footpoints of the 
dark filamentary features of the AFS. In addition, for much of the area covered 
with granulation, the $\chi^2$ goodness-of-fit statistics is low. Good CM 
inversions correspond in general to regions, where the contrast profiles exhibit 
enhanced contrast as is evident in Fig.~\ref{ABSCON} for the AFS. Only for 
regions with high positive contrasts, that is, the footpoint regions, CM inversions 
fail.

\begin{figure}[t]
\includegraphics[width=\columnwidth]{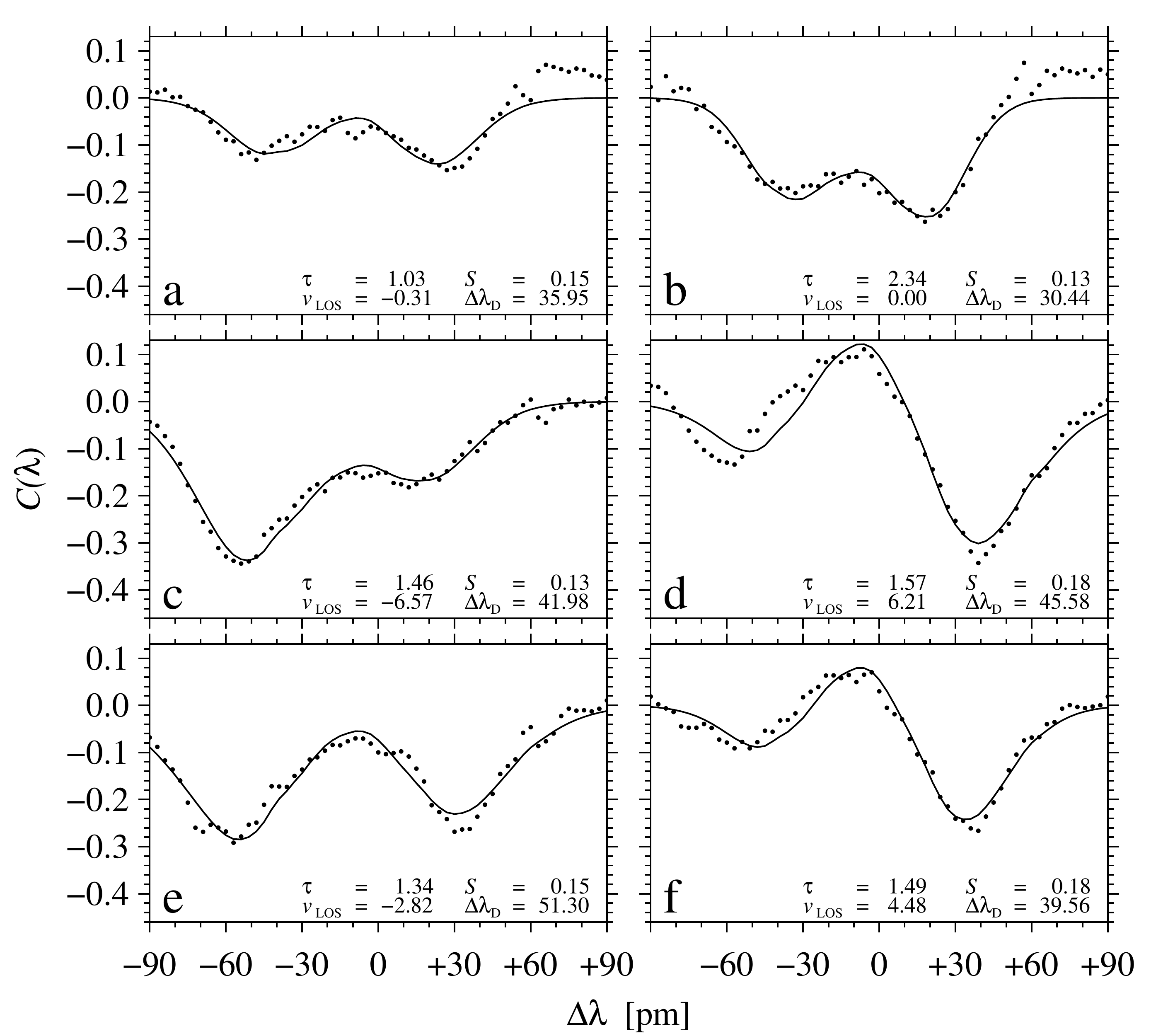}
\caption{Observed contrast profiles $C(\lambda)$ (dotted) and results of CM 
    inversions $C^\prime(\lambda)$ (solid). The CM parameters (omitting the 
    commonly used units) of the inversions are given in the lower right corner 
    of each panel. The alphabetical labels correspond to the locations marked 
    in Fig.~\ref{FIG01}.}
\label{FIG_PLOT_CONTRAST_FIT}
\end{figure}

Typical examples of observed $C(\lambda)$ and fitted contrast profiles 
$C^\prime(\lambda)$ are shown in Fig.~\ref{FIG_PLOT_CONTRAST_FIT} to illustrate 
the quality of the CM inversions (positions of the profiles marked in Fig.~\ref{FIG01}). 
The contrast profiles labeled `a' and `b' have 
low and high values of the optical thickness $\tau_0$, respectively. Profile a 
is taken from the quiet-Sun region with an optical thickness $\tau_{0} \approx 
1$, while profile b belongs to a prominent dark filamentary feature with 
$\tau_{0} \approx 2.3$. In both cases, the velocity $v_\mathrm{LOS}$ is close to 
zero. Contrast profiles c and d correspond to a dark filamentary feature and 
a location near the right H$\alpha$ brightening, respectively. They differ 
mainly in the direction of the velocity $v_\mathrm{LOS}$ and source function 
$S$. The other two CM fit parameters are very similar. The final two examples
e and f are related to 
counter-streaming in a dark filamentary feature (see 
Sect.~\ref{SEC3.3}) exhibiting blue- and redshifts, respectively. 
Counter-streaming is evident in time-lapse movies of H$\alpha$ line-core images. 
The blueshifted profile is characterized by a very high value of 
$\Delta\lambda_D = 51.3$~pm and a relatively low value of $\tau_{0}=1.3$. 

\begin{figure}[t]
\includegraphics[width=\columnwidth]{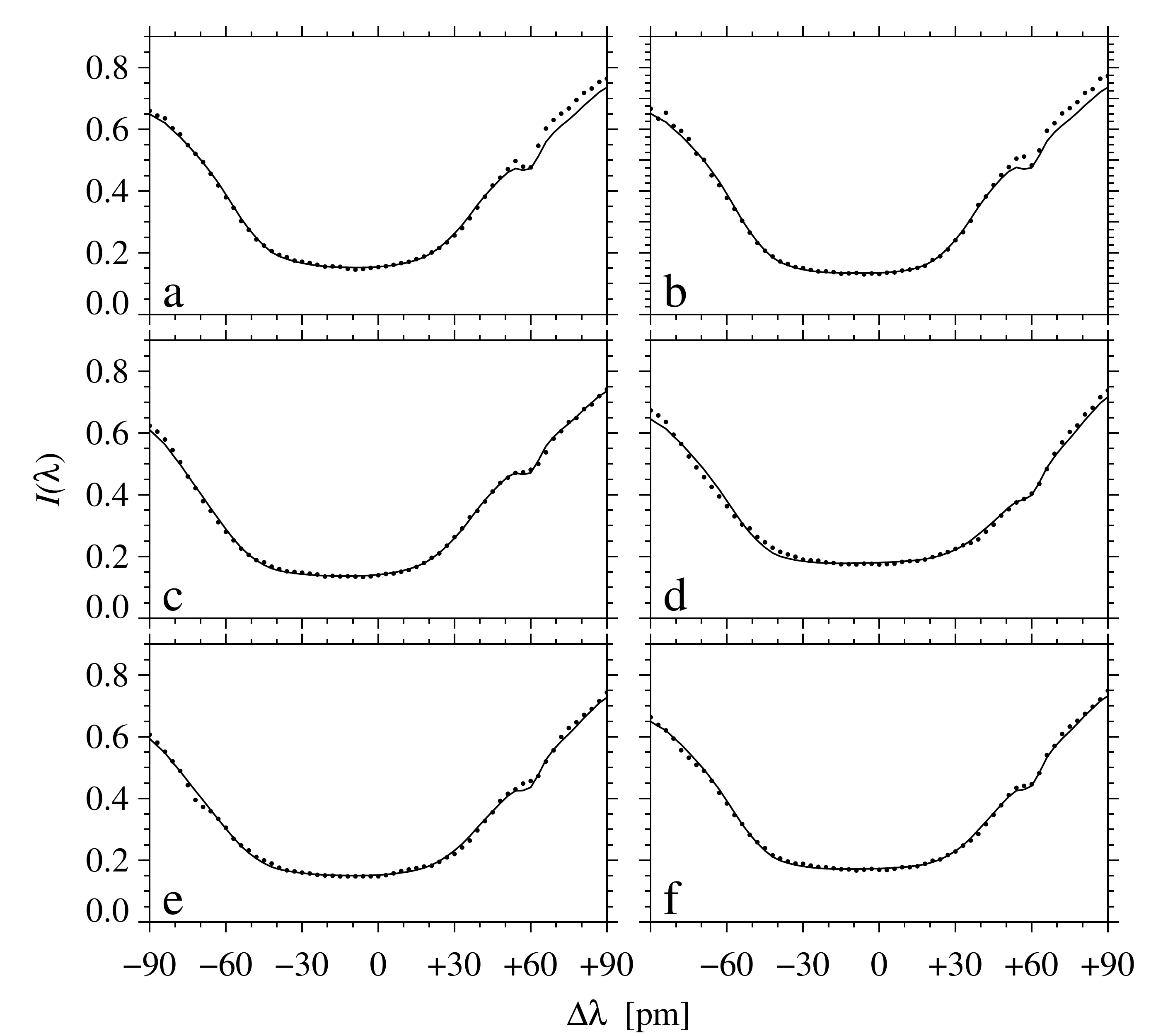}
\caption{Observed normalized H$\alpha$ intensity profiles 
    $I(\lambda)$ (dotted) and converted results of CM inversions 
    $I^\prime(\lambda)$ (solid). The profiles correspond to the contrast 
    profiles $C(\lambda)$ in Fig.~\ref{FIG_PLOT_CONTRAST_FIT}.}
\label{FIG_PLOT_CONTRAST_FIT_REFEREE}
\end{figure}

A fit of an observed H$\alpha$ profile can be derived by 
solving Eqn.~\ref{eq1} for $I(\lambda)$ using the  respective fitted contrast 
profile $C^\prime(\lambda)$ and the quiet-Sun profile $I_0(\lambda)$. The 
H$\alpha$ profiles corresponding to Fig.~\ref{FIG_PLOT_CONTRAST_FIT} are shown 
in Fig.~\ref{FIG_PLOT_CONTRAST_FIT_REFEREE}. Generally, observed and fitted 
H$\alpha$ profiles are in very good agreement. Only for H$\alpha$ profiles with 
lower line depths fits and observations significantly deviate, i.e., in regions 
with high $\chi^2$ values in Fig.~\ref{CHI2}. This test lends additional 
credibility to the results of the CM inversions.

 \subsection{Cluster analysis}\label{SEC5.3}

The four fit parameters of the CM inversion are available for 
all pixels within the FOV, where the least-squares minimization algorithm 
converged and where the $\chi^2$ values are acceptable. However, the presence of 
different features (arch filaments, bright footpoint regions, quiet Sun, etc.) 
within the FOV raises the question if these populations can be characterized by 
distinct CM parameters. Cluster analysis \citep{Everitt2011} is a powerful tool 
to identify populations in an $n$-dimensional parameter space and to locate them 
in the FOV. These populations might be hidden in histograms of CM parameters 
(see Sect.~\ref{SEC3.5}), which can only provide hints of their presence due to 
a particular shape of the contribution. Since $n=4$ for the CM parameter space, 
cluster analysis only delivers meaningful results for less than four 
populations, where two is the most likely number of clusters given the 
histograms of CM parameters in Sect.~\ref{SEC3.5}). The cluster finding 
algorithm, a built-in function of the Interactive Data Language (IDL), maximizes 
the Euclidean distance of the cluster centers while minimizing the inner-cluster 
distances in the CM parameter space. The algorithm does not deliver the number 
of clusters. Thus, a priori knowledge is required about the number of clusters.

%
%

\section{Results}\label{SEC3}

The broadband image in Fig.~\ref{FIG01} contains two micro-pores 
\citep[see][]{RouppevanderVoort2006} within the quiet Sun, which are indications 
of a small EFR close to disk center. As seen in MDI magnetograms, a small, 
compact negative-polarity feature emerges first at 8:00~UT on 2008 August~6 (not 
shown here). In the next 96-minute magnetogram (see Fig.~\ref{BIPOLE}), the 
positive polarity emerges, more extended and with a faint halo-like structure. 
The positive-polarity patch splits in several kernels while the bipole evolves. 
The more stable negative-polarity region hosts the two micro-pores. 
Tracking the EFR was possible until 23:58~UT on 2008 August~7, 
when a data gap of 14~hours prevented us from reliably identifying the EFR, 
that is, only (isolated) small-scale magnetic features were present at the 
predicted location. Therefore, we conclude that the lifetime of the EFR was at 
least 40~hours and at maximum 54~hours. Even though the EFR is small, it 
follows the Hale-Nicholson law \citep{Tlatov2010} and possesses the typical 
magnetic configuration for the $23^\mathrm{rd}$ solar cycle. The dipole emerged 
around 24~hours before the observations were taken, and the total magnetic flux 
increased until the beginning of our observations (see Sect.~\ref{SEC3.1}). 
Although the total magnetic flux in the EFR remained constant for some time 
after our observations, we use the term EFR throughout the article because of 
its bipolar structure and the presence of the small AFS.

\begin{figure*}[t!]
\includegraphics[width=\textwidth]{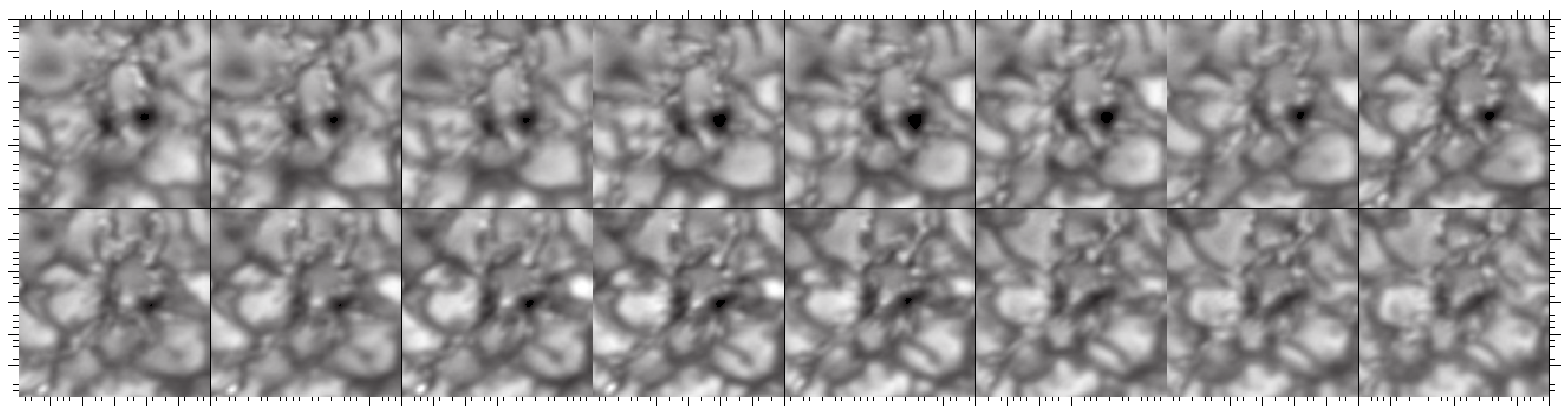}
\caption{Temporal evolution of micro-pores shown at 34-second intervals starting
    at 08:07~UT (\textit{top-left to bottom-right}). The ROI is $6\arcsec 
    \times 6\arcsec$ as indicated by the white square in Fig.~\ref{FIG01}. 
    Major tick marks are separated by one arcsecond. All 
    images are from the reconstructed broadband channel and are displayed in 
    the same intensity range of 0.75\,--\,1.35\,$I_0$.}
\label{FIG10}
\end{figure*}


\subsection{Temporal evolution of micro-pores}\label{SEC3.1}

A region-of-interest (ROI) with a size of $6\arcsec \times 6\arcsec$ centered on 
the micro-pores (white square in Fig.~\ref{FIG01}) is chosen to illustrate the 
temporal evolution of these structures in Fig.~\ref{FIG10}. The time sequence 
starts at 08:07~UT, the cadence is $\Delta t = 34$~s, and the total duration of 
the time sequence is about $\Delta T \sim$10~min.

\begin{figure}[t]
\includegraphics[width=\columnwidth]{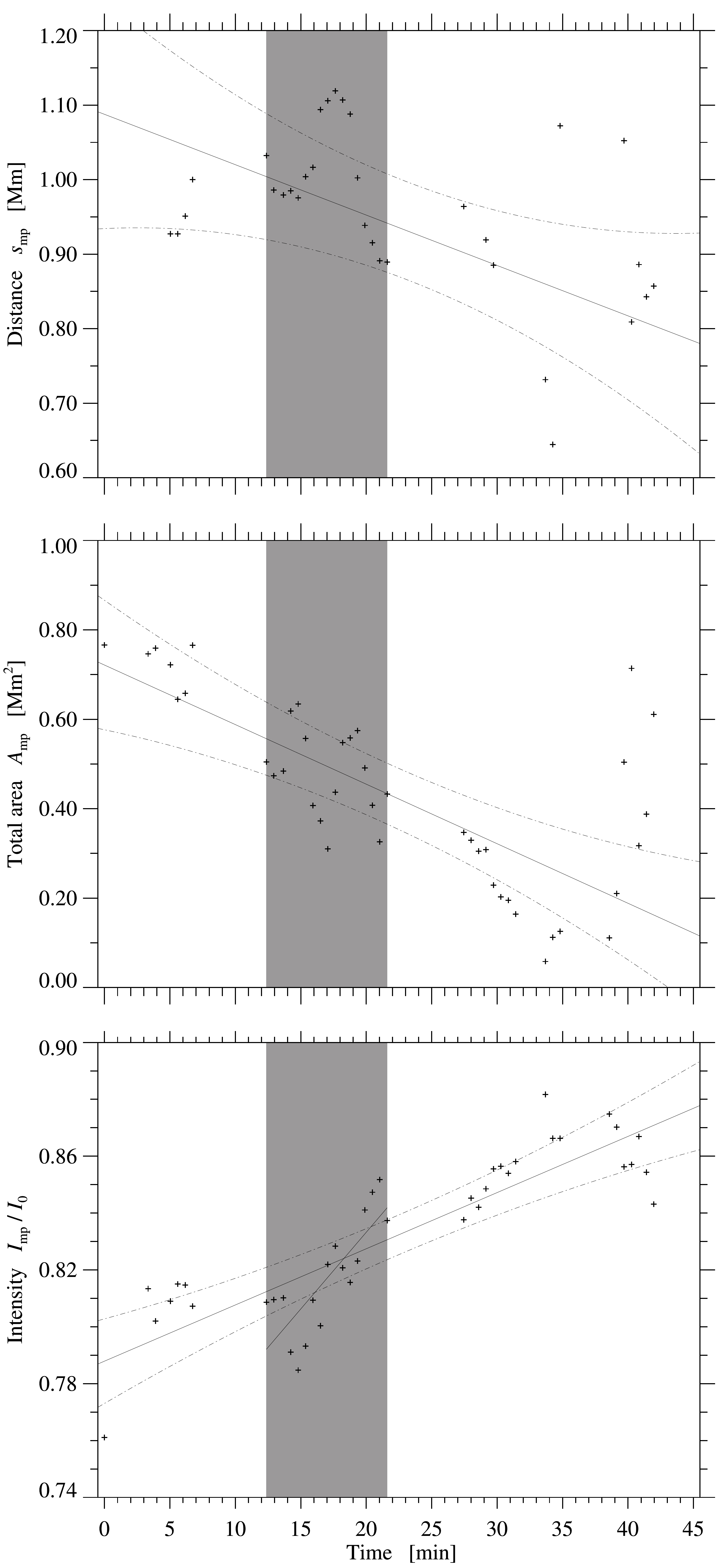}
\caption{Chronological sequence of parameters describing the decay process of
    the micro-pores starting at 07:54~UT, i.e., center-to-center distance of the
    micro-pores $s_\mathrm{mp}$, total area $A_\mathrm{mp}$, and average
    intensity $I_\mathrm{mp} / I_0$ (\textit{top to bottom}). The gray
    background indicates the 17 sequences with the best seeing conditions. The 
    solid lines are linear fits to the parameters, and the dash-dotted lines 
    represent the 3$\sigma$-error margins.}
\label{FIG_MICRO_PORES_BLOB_ALL}
\end{figure}

Initially, two micro-pores are present with diameters of less than $1\arcsec$, 
which evolve with time in intensity, size, and shape. The right micro-pore is 
circular, whereas the left micro-pore exhibits some star-like extrusions. The 
intensity of both micro-pores gradually increases, and at the same time the area 
decreases. As well as the term `micro-pore', \citet{RouppevanderVoort2005} 
introduced the nomenclature `ribbon' and `flower' for elongated and more 
circular magnetic structures, respectively, to capture their morphology at 
sub-arcsecond scales. The transition from flower to ribbon begins for the left 
micro-pore early-on in the time series, whereas the right micro-pore maintains 
its perceptual structure for about three minutes and then makes the transition 
from flower to ribbon. Small-scale brigthenings and signatures of abnormal 
granulation \citep{DeBoer1992} are present everywhere in the vicinity of the 
micro-pores (e.g., near the coordinates (30\arcsec, 30\arcsec) and (29\arcsec, 
21\arcsec) in Fig.~\ref{FIG01}).

\begin{figure}[t]
\includegraphics[width=\columnwidth]{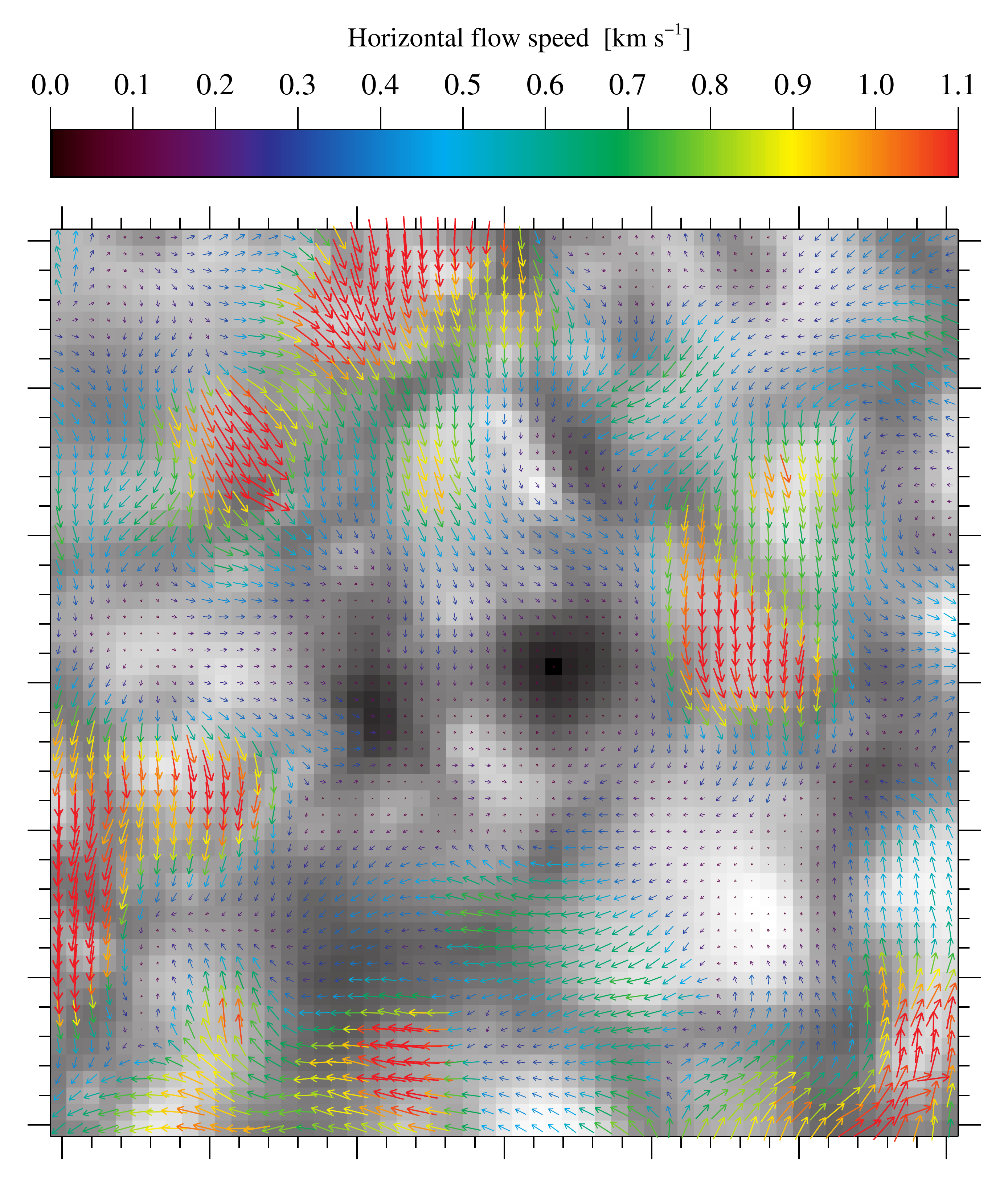}
\caption{Horizontal proper motions in the vicinity of the two micro-pores.
    Color-coded arrows indicate magnitude and direction of horizontal flows for
    each pixel of the first broad band image shown in Fig.~\ref{FIG10}, an
    enlarged version of which serves as the background. Major tick marks are 
    separated by one arcsecond. The FOV $6\arcsec \times 6\arcsec$
    as indicated by the white square in Fig.~\ref{FIG01}.}
\label{FIG_LCT}
\end{figure}

The ten-minute time series in Fig.~\ref{FIG10} displays the two micro-pores 
under the best seeing conditions. However, the other three time series also 
contain moments of good seeing, thus allowing us to study the evolution of the 
micro-pores more quantitatively over a period of about 40 minutes. Smoothing 
using anisotropic diffusion \citep{Perona1990}, intensity thresholding 
($<0.9\,I_0$), and standard tools for `blob analysis' \citep{Fanning2011} yield 
the area of the micro-pores, their intensity distribution, and their center 
coordinates, which in turn provide the distance between the micro-pores. A more 
detailed description of this procedure is given in \citet{Verma2014}. The 
temporal evolution of the total area $A_\mathrm{mp}$ of the micro-pores, their 
average intensity $I_\mathrm{mp}$ given in terms of the quiet-Sun intensity 
$I_0$, and the center-to-center distance $s_\mathrm{mp}$ are presented in 
Fig.~\ref{FIG_MICRO_PORES_BLOB_ALL}. Because the micro-pores evolve in a dynamic 
environment, being constantly jostled around by granules, they change shape and 
sometimes even seem to merge or split (mediocre seeing also affects the feature 
identification). Therefore, only data points are included for the distance, 
where the two micro-pores can be clearly identified. In addition, some restored 
broadband images were omitted because of mediocre seeing conditions.

All aforementioned parameters indicate the photometric decay of the system of 
micro-pores; the total magnetic flux of the EFR, however, remains constant but 
its distribution within the region changes. At the beginning of the sequence the 
micro-pores are about 1.1~Mm apart, have a total area of around 0.8~Mm$^2$, and 
their average intensity is just $0.75\,I_0$. Linear least-squares fits establish 
that the two micro-pores approach each other with a speed of $ds_\mathrm{mp} / 
dt = -0.11$~km~s$^{-1}$, shrink at a rate of $dA_\mathrm{mp} / dt = 
-0.22$~km$^2$~s$^{-1}$, and approach quiet-Sun intensity levels at 
$dI_\mathrm{mp} / dt = 0.033\,I_0$~s$^{-1}$. 

Measuring the magnetic flux and thus the growth and decay rates is hampered by 
the low cadence (96~min), the varying noise level as a function of the 
heliocentric angle, the geometric foreshortening of the pixel area, the angle of 
the magnetic field lines (assumed to be perpendicular to the surface) with the 
line-of-sight (LOS), and the small number of pixels having a flux density of 
more than 20~G. Therefore, we conservatively report only the total unsigned 
magnetic flux of $2.2 \times 10^{20}$~Mx (the negative magnetic flux is $1.2 
\times 10^{20}$~Mx) at 08:00~UT on 2008 August~7, which attains a maximum right 
at the time of the high-resolution GFPI observations. The flux contained in the 
EFR is slightly larger than the upper limit for ephemeral regions 
\citep[see][]{vanDrielGesztelyi2015}, that is, much too small to form larger pores 
or even sunspots. Even though the time resolution is too coarse to provide a 
growth rate for the magnetic field, our measurements agree with a rapid rise 
time (already significant changes between the first two magnetograms) and a much 
slower decay rate (total unsigned flux of $1.3 \times 10^{20}$~Mx at 22:24~UT on 
2008 August~7). However, we cannot use the MDI continuum images to determine the 
lifetime of the micro-pores because their size is only a small fraction of an 
MDI pixel. 

The $3\sigma$-errors given in Fig.~\ref{FIG_MICRO_PORES_BLOB_ALL} demonstrate 
that a linear model for the decay process of the micro-pores is in general 
appropriate -- at least for the time interval under study. However, there are 
some deviations, in particular during the time with the best seeing (highlighted 
gray area in Fig.~\ref{FIG_MICRO_PORES_BLOB_ALL}), where a more detailed 
inspection is warranted. The two micro-pores start to separate, while becoming 
darker. During this period the left micro-pore almost vanishes while the one on 
the right changes its shape from roundish to more elongated. Toward the end of 
the sequence shown in Fig.~\ref{FIG10}, the left micro-pore slowly recovers. A 
linear least-squares fit for just this time period  yields an increased 
intensity gradient of $dI_\mathrm{mp} / dt = 0.045\,I_0$~s$^{-1}$. The increase 
in total area toward the end of the sequence is caused by newly forming and 
merging micro-pores. A slight decrease in intensity accompanies this emergence 
process, as expected for a small-scale magnetic flux kernel with increasing 
magnetic flux. Taking the temporal derivatives $ds_\mathrm{mp} / dt$, 
$dA_\mathrm{mp} / dt$, and $dI_\mathrm{mp} / dt$ at face value leads to the 
conclusion that the decaying micro-pores will vanish before merging.


\subsection{Horizontal flow field around micro-pores}\label{SEC3.2}

Magnetic field concentration are known to influences 
horizontal proper motions. We used LCT \citep{Verma2011} to investigate if even 
micro-pores potentially impact the surrounding plasma motions. The mean flow 
speed is  $\bar{v} = 0.59 \pm 0.41$~km~s$^{-1}$  across the entire FOV and 
$\bar{v} = 0.49 \pm 0.31$~km~s$^{-1}$ in the ROI shown in Fig.~\ref{FIG_LCT}. In 
both cases, the standard deviation refers to the variation within the observed 
field rather than to a formal error estimate. The corresponding 10$^\mathrm{th}$ 
percentiles of the speed distributions are $v_{10} = 1.15$ and 0.93~km~s$^{-1}$, 
respectively. These values are slightly higher than those provided in Fig.~7 of 
\citet{Verma2013}. However, considering the shorter averaging time of $\Delta T 
\approx 10$~min as compared to $\Delta T = 60$~min for Hinode G-band images, the 
LCT results are in good agreement with the values presented by \citet{Verma2013}.

Neither inflows nor outflows surround the two micro-pores. The flow field is still 
dominated by the jostling motion of the solar granulation. As already noted in 
\citet{Verma2011}, much longer averaging times are needed to clearly detect 
persistent flows. However, the flow speed is significantly 
reduced in the immediate surroundings of the micro-pores ($\pm$1\arcsec). 
Detecting (micro-) pores at spatial scales of one arcsecond or below is a 
formidable task. In their extensive statistical study based on Hinode data, 
\citet{Verma2014} consequently required an area of at least 0.8~Mm$^2$ for pores 
to be included in the sample, which would exclude the two micro-pores depicted in 
Fig.~\ref{FIG_LCT}. Therefore, this case study provides additional information 
for the smallest features that can still be considered to be pores.


\subsection{Chromospheric fine structure and Doppler velocity}\label{SEC3.3}

\begin{figure}[t]
\includegraphics[width=\columnwidth]{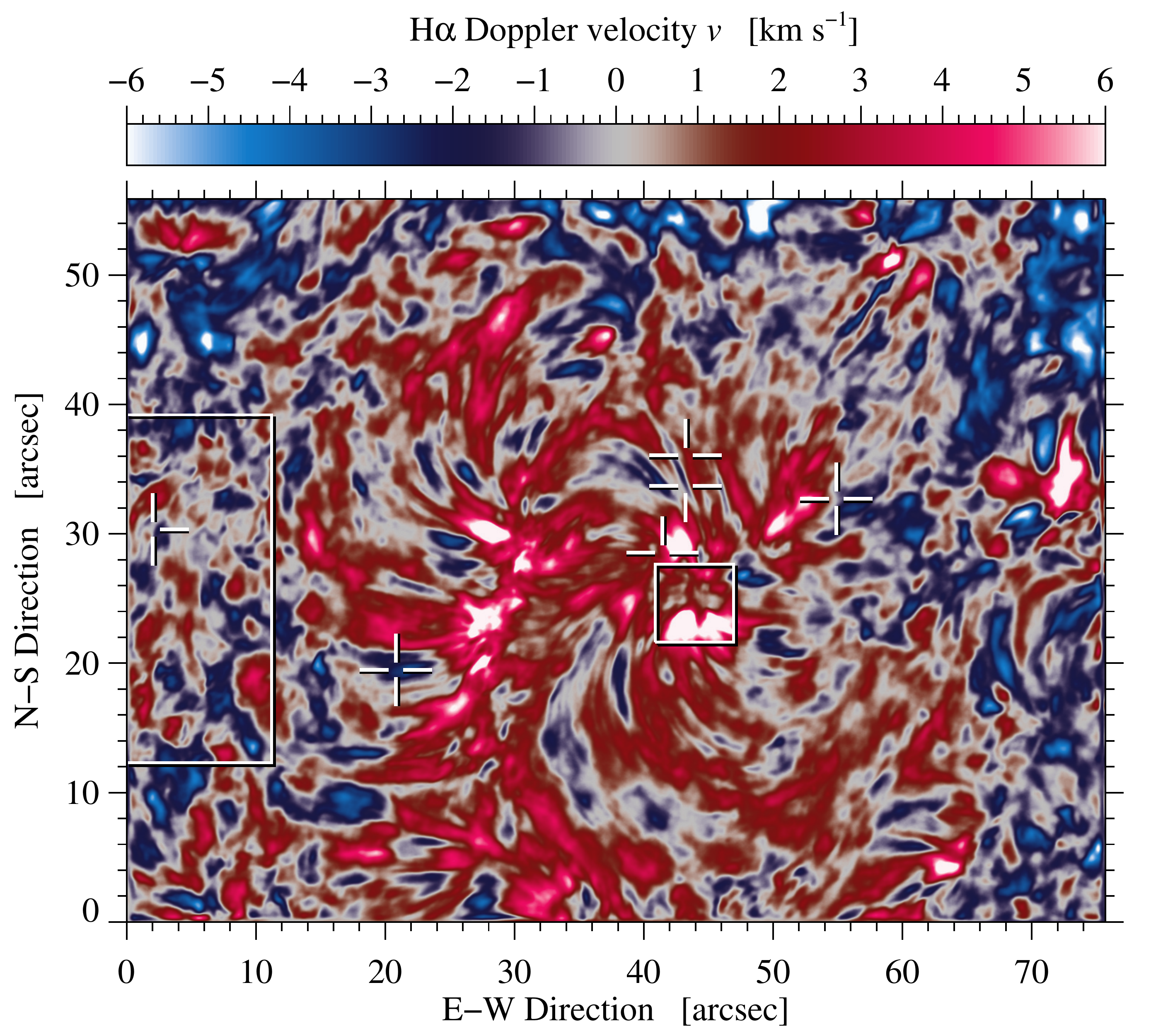}
\caption{Chromospheric Doppler velocity map corresponding to Fig.~\ref{FIG01}
    derived with the center-of-gravity method. Blue and red colors represent
    up- and downflows, respectively.}
\label{FIG05}
\end{figure}

Both the H$\alpha$ line-core intensity (bottom panel in Fig.~\ref{FIG01}) and 
the H$\alpha$ Doppler velocity maps (Fig.~\ref{FIG05}) reveal the rich fine 
structure of the solar chromosphere. The center-of-gravity method 
\citep{Schmidt1999} is very efficient and robust in retrieving the H$\alpha$ 
Doppler velocities. Repeating the white rectangle and square already drawn in 
Fig.~\ref{FIG01} facilitates easy comparison with the photospheric observations 
-- in particular, the rectangle validates the choice of the quiet-Sun region as 
the Doppler velocity reference, because this region is free of any chromospheric 
filamentary structure and does not exhibit any peculiar chromospheric flows; and 
the square illustrates the association of the micro-pores with H$\alpha$ 
brightenings and strong downflows in excess of 6~km~s$^{-1}$ near the footpoints 
of small-scale H$\alpha$ loops belonging to the AFS. However, the Doppler 
velocities are close to zero at the exact location of the micro-pores.

The AFS consists of two regions with H$\alpha$ line-core brightenings. Some 
small-scale loops connect these areas containing the bipolar EFR. Other loops 
rooted in the bright patches connect to the  outside of the EFR. Some bright 
fibrils are interspersed among the dark filaments. The width of the loops is 
just a few seconds of arc, and their length is about 6\arcsec\,--\,8\arcsec. 
Typical aspect ratios for the dark filaments are 1\tsp:\,4 to 1\tsp:\,5. The 
morphology of the micro-pores, the EFR, and the small AFS closely resembles 
structures reported in other studies \citep[e.g.,][]{Strous1996, Tziotziou2003, 
WedemeyerBoehm2009, Zuccarello2009b}. The cross-shaped markers labeled 
a\,--\,f identify the locations of some exemplary H$\alpha$ contrast profiles, 
for example, labels c  and d refer to a dark arch filament and a footpoint, 
respectively. Properties and parameters derived from CM inversions of the 
contrast profiles are discussed in Sect.~\ref{SEC3.5}.

\begin{figure}[t]
\includegraphics[width=\columnwidth]{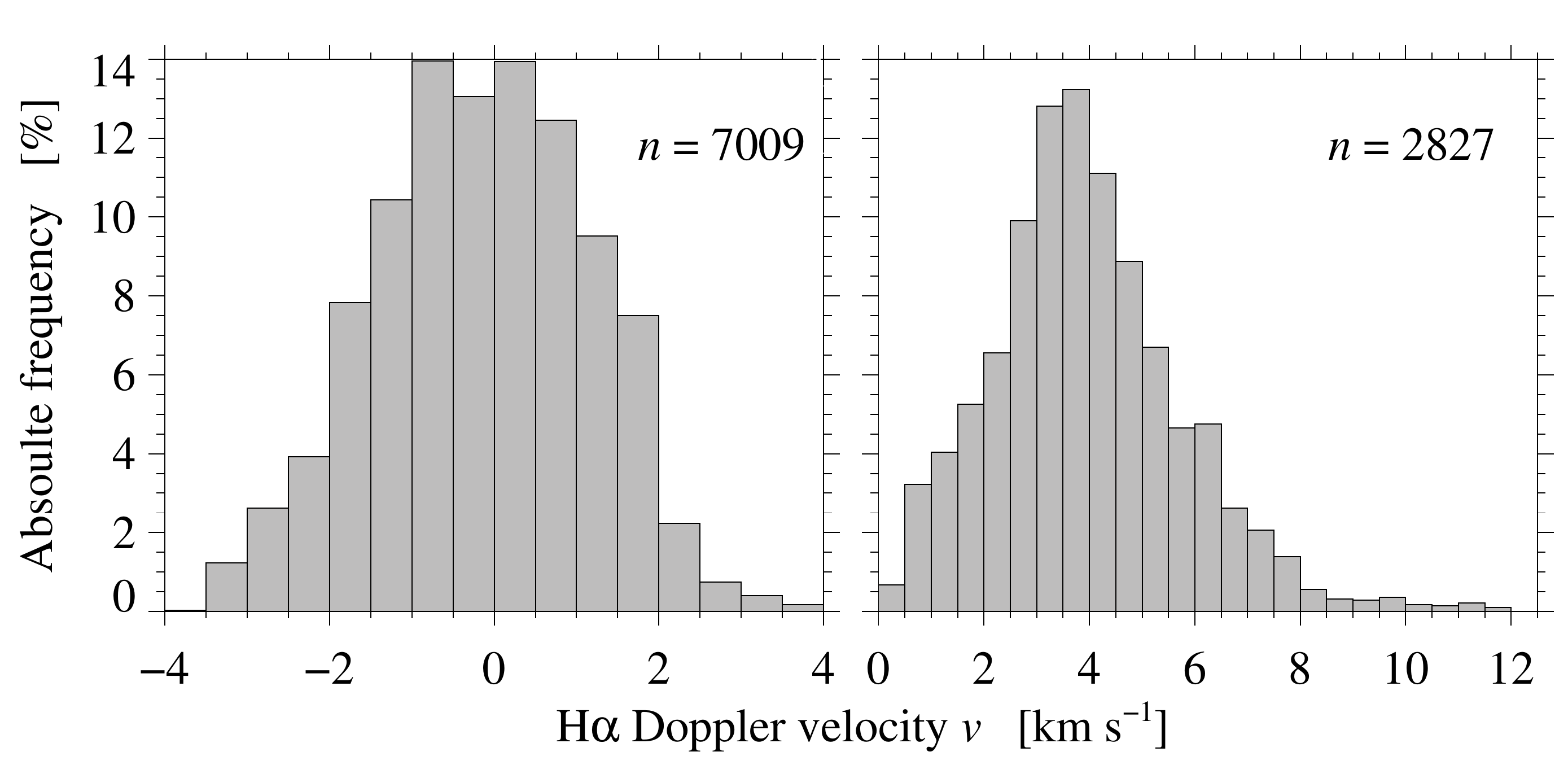}
\caption{Normalized histograms of the chromospheric Doppler 
    velocities $v_d$ for dark filamentary features (\textit{left}) and $v_b$ 
    for bright footpoints (\textit{right}) of the small AFS in 
    Figs.~\ref{FIG01} and \ref{FIG05}. The variable $n$ indicates the number of 
    H$\alpha$ line profiles on which the distributions are based.}
\label{FIG11}
\end{figure}

H$\alpha$ Doppler velocities $v$ are computed for the dark filaments of the AFS 
and the bright footpoint areas. Intensity and size thresholds in combination 
with morphological image processing of the H$\alpha$ line-core intensity maps 
yield binary masks of bright and dark features. Indexing of pixels belonging to 
these areas facilitates computing histograms for dark 
filamentary features ($n = 7009$ pixels) and bright footpoints  ($n = 2827$ 
pixels) shown in the left and right panels of Fig.~\ref{FIG11}, respectively. 
The histogram of the dark filamentary features is 
characterized by a median value of $v_{d,\mathrm{med}} = -0.10$~km~s$^{-1}$, a 
mean value of $\bar{v}_d = -0.13$~km~s$^{-1}$, and a standard deviation of 
$\sigma_{d} = 1.31$~km~s$^{-1}$. The corresponding values for the bright 
footpoints are $v_{b,\mathrm{med}} = 3.77$~km~s$^{-1}$, $\bar{v}_b = 
3.97$~km~s$^{-1}$, and $\sigma_{b} = 1.86$~km~s$^{-1}$.

Both up- and downflows occur in the dark filaments with a small imbalance 
toward blueshifts. Notwithstanding, the bright footpoint regions almost 
exclusively harbor downflows with velocities as high as 12~km~s$^{-1}$. 
Interestingly, blue- and redshifted H$\alpha$ profiles appear in close proximity 
along the loops (see labels e and f), which can be interpreted as up- and 
downflows. However, counter-streaming and even rotation along the filament axis 
are possibilities depending on the specific three-dimensional loop topology. The 
typical mean upflow velocity at the loop tops is $\bar{v}_{lt} = 
-1.13$~km~s$^{-1}$, and extreme values approach $v_{lt,\mathrm{max}} = 
-3$~km~s$^{-1}$. In general, the ratio of the area between up- and downflowing 
cloud material is $A_\mathrm{up}/A_\mathrm{down} = 0.22$.


\subsection{Cloud model inversions}\label{SEC3.4}

\begin{table}[t]
\caption{Mean value and standard deviation of CM parameters for the quiet Sun, 
    the dark filamentary features, and all fitted contrast profiles.}
\label{tab:cm_results}
\begin{center}
\setlength{\tabcolsep}{2.5pt}
\small
\begin{tabular}{llccc}
\hline\hline
\rule[-2mm]{0mm}{6mm} &                   & quiet Sun                 & dark filaments       & all profiles\\
\hline
$S$                   &                   & $ 0.14 \pm 0.07$          & $0.12 \pm 0.02$      & $ 0.15 \pm  0.05$\rule{0mm}{4mm}\\
$\Delta\lambda_D$     & [pm]              & $30.44 \pm 10.36$         & $41.62 \pm 7.17$\phn & $33.96 \pm 10.74$\\
$\tau_0$              &                   & $ 0.44 \pm 0.30$          & $1.19 \pm 0.45$      & $ 0.85 \pm  0.53$\\
$v_\mathrm{LOS}$      & [km~s$^{-1}$]\phn & \phn$-1.41 \pm 10.16$\phm & $-0.94 \pm 3.84$\phm & $ 0.41 \pm  8.61$\rule[-2mm]{0mm}{4mm}\\ 
\hline
\end{tabular}
\end{center}
\end{table}

\begin{table}[t]
\caption{Gaussian and log-normal fit parameters of the normalized  
    histograms of the CM parameters shown in 
    Fig.~\ref{FIG08}.}
\label{TAB02}
\begin{center}
\small
\begin{tabular}{llccccc}
\hline\hline
\rule[-2mm]{0mm}{6mm}   &               & $\mu_1$    & $\sigma_1$ & $\mu_2$     
& $\sigma_2$ & $P_1$ \\ \hline
$S$                     &               & $0.14$     & $0.02$     & $0.18$      
& $0.01$     & $0.83$ \rule{0mm}{4mm}\\
$\Delta\lambda_D$       & [pm]          &$25.92$\phn & $6.81$     & $42.09$\phn 
& $6.35$     & $0.54$ \\
$\tau_{0}{^\dagger}$        &               & $0.17$     & $3.5$     & $1.00$      
& $2.60$     & $0.50$ \\
$v_\mathrm{LOS}$        & [km~s$^{-1}$] & $0.61$     & $9.77$     & $0.39$      
& $3.45$     & $0.76$ \rule[-2mm]{0mm}{4mm}\\ 
\hline
\end{tabular}
\parbox{78mm}{\vspace*{2mm}
\footnotesize{$^{\dagger}$Two Gaussians are fitted to the normalized  
    histograms of the CM fit parameters with the exception 
    of $\tau_0$, where the fit consists of a two log-normal distributions.}}
\end{center}
\end{table}

The rms-deviation between observed and inverted contrast profiles is taken as an 
additional criterion to select only the best-matched profiles (60\% of the 
entire FOV) as shown in Fig.~\ref{FIG09}, where mediocre fits are excluded and 
indicated as light gray areas. Rejected regions include much of the quiet Sun 
and the micro-pores including the surrounding regions with H$\alpha$ line-core 
brightenings. In all cases, the underlying assumption of CM inversions is not 
valid, that is, the premise of cool absorbing plasma suspended by the magnetic 
field above the photosphere does not apply. Mean values and standard deviations 
of the CM fit parameters are summarized in Table~\ref{tab:cm_results} for the 
quiet-Sun region indicated by the white rectangles in Figs.~\ref{FIG01} and 
\ref{FIG05}, the dark filamentary features, and all fitted profiles. The dark 
filamentary features represent just 2\% of all contrast profiles in the image.

In all three cases, the source function $S$ is virtually the same, except for the 
lower standard deviation of the dark filamentary features. The Doppler width 
$\Delta\lambda$ and the optical thickness $\tau_0$ are significantly larger in 
the dark filamentary features as compared to quiet-Sun regions. The dark 
features show on average a small upflow similar to the quiet Sun. The quiet-Sun 
LOS velocity $v_\mathrm{LOS}$ is not zero like the H$\alpha$ Doppler velocity 
$v$, which is likely a selection effect because CM inversions are not available 
for all pixels of the reference region. The LOS velocity $v_\mathrm{LOS}$ for 
the entire FOV is slightly redshifted because of the downflows encountered in 
the EFR and AFS. The small standard deviation of all CM fit parameters for the 
dark filamentary features indicates that they belong to a specific class of 
absorption features.

\begin{table*}[t]
\caption{Results of cluster analysis for the CM parameters
    depicted in in Figs.~\ref{FIG08} and \ref{FIG09}.}
\label{TAB03}
\begin{center}
\fontsize{8}{10}\selectfont
\begin{tabular}{llcccccccc}
\hline\hline
\multicolumn{2}{l}{2 populations}\rule[-2mm]{0mm}{6mm}               & $w_1$       & $w_2$       & $w_3$ & $w_4$  & $\mu_1\pm\sigma_1$              & $\mu_2\pm\sigma_2$            & $\mu_3\pm\sigma_3$              & $\mu_4\pm\sigma_4$                  \\ \hline
$S$                       &               & $0.15$      & $0.15$      &                &             & $0.13 \pm  0.04$     & $0.15 \pm 0.04$    &                      &                          \rule{0mm}{4mm}\\
$\Delta\lambda_D$         & [pm]          &$16.00$\phn  &$47.00$\phn  &                &             &$23.96 \pm  4.98$\phn &$41.54 \pm 5.57$\phn&                      &                          \\
$\tau_0$        &               & $0.31$      & $0.33$      &            &             & $0.66 \pm  0.45$     & $1.02 \pm 0.52$    &                      &                          \\
$v_\mathrm{LOS}$          & [km~s$^{-1}$] &$-7.03$\phn  &$-6.29$\phn  &            &             & \phn$0.67 \pm 10.58$ & $0.19 \pm 5.84$    &                      &                          \rule[-2mm]{0mm}{4mm}\\                        
\hline
\multicolumn{2}{l}{Cluster fraction}\rule[-2mm]{0mm}{6mm} &$c_1 = 43.4$\% & $c_2 = 56.6$\% & & & & & & \\ 
\hline
\multicolumn{10}{l}{3 populations}\rule[-2mm]{0mm}{6mm} \\ \hline
$S$                       &               &  $0.14$      & $0.15$      & $0.15$    &             &  $0.13 \pm 0.05$     & $0.13 \pm 0.04$    & $0.15 \pm 0.04$      &                          \rule{0mm}{4mm}\\
$\Delta\lambda_D$         & [pm]          & $20.18$\phn  &$16.80$\phn  &$47.50$    &             &$24.99 \pm 5.15$\phn  &$25.37 \pm 6.19$\phn&$42.30 \pm 6.52$\phn  &                          \\
$\tau_0$        &               &  $0.36$      & $0.34$      & $0.32$    &             &  $0.56 \pm 0.39$     & $0.72 \pm 0.48$    & $1.04 \pm 0.52$      &                          \\
$v_\mathrm{LOS}$          & [km~s$^{-1}$] &$-16.04$\phnn  &$10.09$\phn  &$-6.14$\phn&             &$-11.18 \pm 5.34$\phnn & $7.79 \pm 5.88$    &$-0.71 \pm 4.78$\phn  &                          \rule[-2mm]{0mm}{4mm}\\ 
\hline
\multicolumn{2}{l}{Cluster fraction}\rule[-2mm]{0mm}{6mm} &$c_1 = 16.2$\% & $c_2 = 33.0$\ & $c_3 = 50.8$\% & & & & &  \\ 
\hline
\multicolumn{10}{l}{4 populations}\rule[-2mm]{0mm}{6mm} \\ \hline
$S$                       &              &  $0.14$   & $0.15$     & $0.15$     & $0.16$      &  $0.13 \pm 0.05$       & $0.13 \pm 0.04$      & $0.16 \pm 0.04$       & $0.15 \pm 0.03$           \rule{0mm}{4mm}\\
$\Delta\lambda_D$         & [pm]         & $19.73$\phn  &$16.67$\phn  &$48.55$\phn  &$39.71$\phn  & $23.72 \pm 4.98$\phn &$23.81 \pm 5.47$\phn &$49.43 \pm 4.30$\phn &$37.03 \pm 4.91$\phn         \\
$\tau_0$        &     &  $0.32$ & $0.31$ & $0.30$ & $1.18$    &  $0.50 \pm 0.36$  & $0.67 \pm 0.47$ & $1.06 \pm 0.55$ & $1.00 \pm 0.50$                         \\
$v_\mathrm{LOS}$          & [km~s$^{-1}$] &$-16.38$\phnn &$10.20$\phn &$-5.51$\phn &$-4.75$\phn      &$-12.33 \pm 5.18$\phnn & $8.31 \pm 6.10$ &$-0.41 \pm 4.72$\phn &$-0.62 \pm 5.48$\phn             \rule[-2mm]{0mm}{4mm}\\ 
\hline
\multicolumn{2}{l}{Cluster fraction}\rule[-2mm]{0mm}{6mm} & $c_1 = 43.1$\% & $c_2 = 17.2$\% & $c_3 = 27.3$\% & $c_4 = 12.4$\% & & & & \\ 
\hline
\end{tabular}
\parbox{175mm}{\vspace*{2mm}
\footnotesize{The cluster analysis was performed for $k=2$,  3, and 4 
    populations. The weights $w_i$ refer to the cluster centers. In addition, 
    mean values $\mu_i$ along with the standard deviations $\sigma_i$ were 
    computed for the CM parameters of each cluster. The cluster fractions
    $c_i$ are given with respect to all good CM fits.}}
\end{center}
\end{table*}


\subsection{Properties of the arch filament system}\label{SEC3.5}
 
\begin{SCfigure*}[0.75][t]
\includegraphics[width=0.74\textwidth]{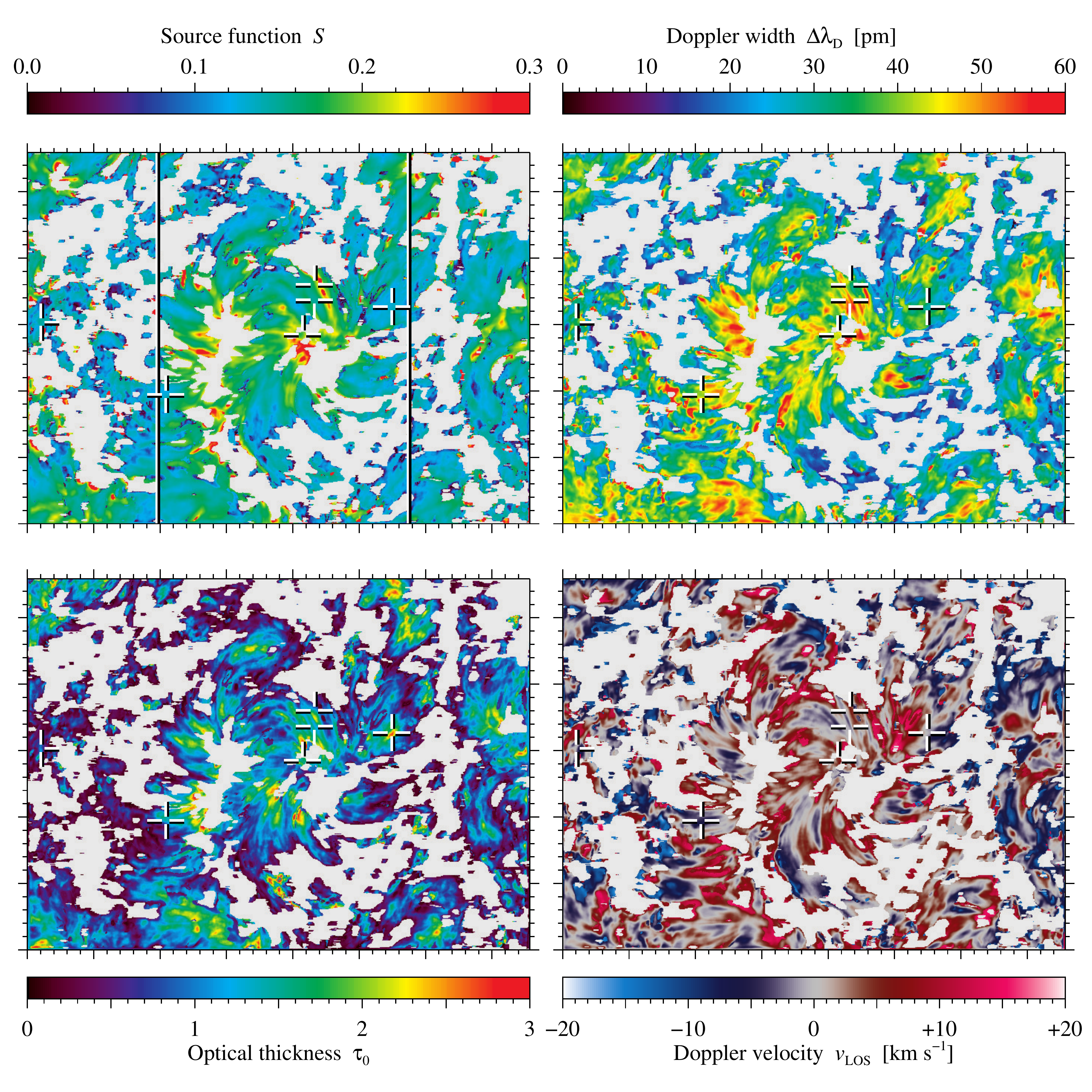}
\hspace*{1mm}
\caption{Maps of CM parameters for the inverted H$\alpha$ contrast profiles 
    corresponding to Fig.~\ref{FIG01}, i.e., source function $S$, Doppler width 
    $\Delta\lambda_D$, optical thickness $\tau_{0}$, and LOS velocity 
    $v_\mathrm{LOS}$ (\textit{top-left to bottom-right}). Light gray areas
    indicate regions, where the CM inversions did not deliver good fits.  
    The two white vertical lines in the upper-left panel 
    enclose the ROI, which was selected to show the evolution of the AFS in 
    Fig.~\ref{evolution_cm}. The FOV is the same as shown in Fig.~\ref{FIG01}.} 
\label{FIG09}
\end{SCfigure*}

The results of the CM inversions are presented in Fig.~\ref{FIG09} for the 
parameters source function $S$, Doppler width $\Delta\lambda_{D}$, optical 
thickness $\tau_0$, and LOS velocity of the cloud material $v_\mathrm{LOS}$, 
which is not the same as the Doppler velocity $v$ obtained with the 
center-of-gravity method. The linear correlation coefficient between these two 
quantities is $\rho (v_\mathrm{LOS},\,v) = 0.80$. The linear regression model 
can be written as $v = c_0 + c_1 \cdot v_\mathrm{LOS}$ with the constants $c_0 = 
0.20$~km~s$^{-1}$ and $c_1 = 0.15$. The linear correlation of $v_\mathrm{LOS}$ 
and $v$ reduces to $\rho_{lt} = 0.73$ and $\rho_{qs} = 0.64$ for specific 
features such as loop tops and the quiet Sun.

The parameters $S$, $\Delta\lambda_{D}$, and $\tau_0$ reach their largest values 
in proximity to the H$\alpha$ line-core brightenings. In regions, which are free 
of any filamentary structure, the values for this set of parameters are much 
lower. The velocity $v_\mathrm{LOS}$ within the EFR and the AFS is predominantly 
directed downwards with the exception of the loop tops of the dark filamentary 
features, where significant upflows arise, which are clearly in excess of the 
H$\alpha$ Doppler velocity $v$ (compare with Fig.~\ref{FIG01}).

\begin{figure}[t]
\includegraphics[width=\columnwidth]{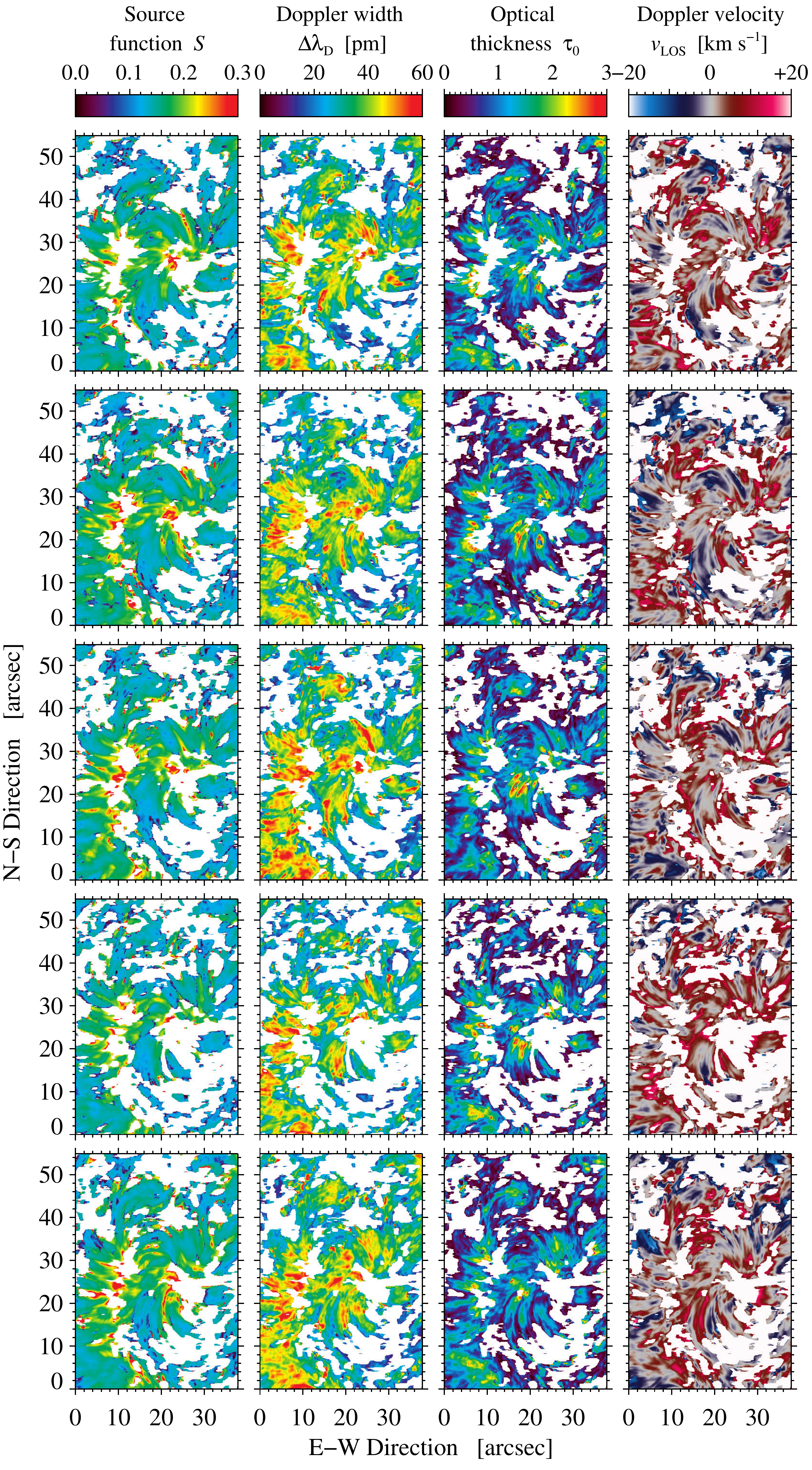}
\caption{Temporal evolution of CM parameters (\textit{top to 
    bottom}) for the ROI containing the EFR and AFS. The maps are depicted at 
    136-second intervals starting at 08:07~UT (see Fig.~\ref{FIG09}) and show
    source function $S$, Doppler width $\Delta\lambda_D$, optical thickness 
    $\tau_{0}$, and LOS velocity $v_\mathrm{LOS}$ (\textit{left to right}).
    The two white vertical lines in the upper-left panel shown in Fig.~\ref{FIG09}
    enclose the FOV of this figure.}
\label{evolution_cm}
\end{figure}

The temporal evolution of selected CM parameter maps is shown 
in Fig.~\ref{evolution_cm} for the central ROI containing the EFR and AFS as 
indicated in the top-left panel of Fig.~\ref{FIG09}. The first row corresponds 
to Fig.~\ref{FIG09} at 08:07~UT, whereas the other rows display maps at 
136-second intervals (every fourth image in Fig.~\ref{FIG10}). The CM inversions 
and the quiet-Sun selection follow the same procedure as for the first CM map. 
The backgrounds of source function $S$ and optical thickness $\tau_0$ remain 
very stable but changes of their extreme values closely track the evolution of 
the dark arch filaments. The Doppler width $\Delta\lambda_{D}$ of the arch 
filaments reaches a maximum in the third map. The velocity $v_\mathrm{LOS}$ at 
the loop tops reveals upflows in the first three maps while in the last two maps 
downflows appear at the same position. This points to the very dynamic nature of 
AFSs. Despite individual active features, in particular the loop tops, on the 
whole the AFS is very stable over the 10-minute time series, which is consistent 
with the results of \citet{Tsiropoula1992}.

\begin{figure}[t]
\includegraphics[width=\columnwidth]{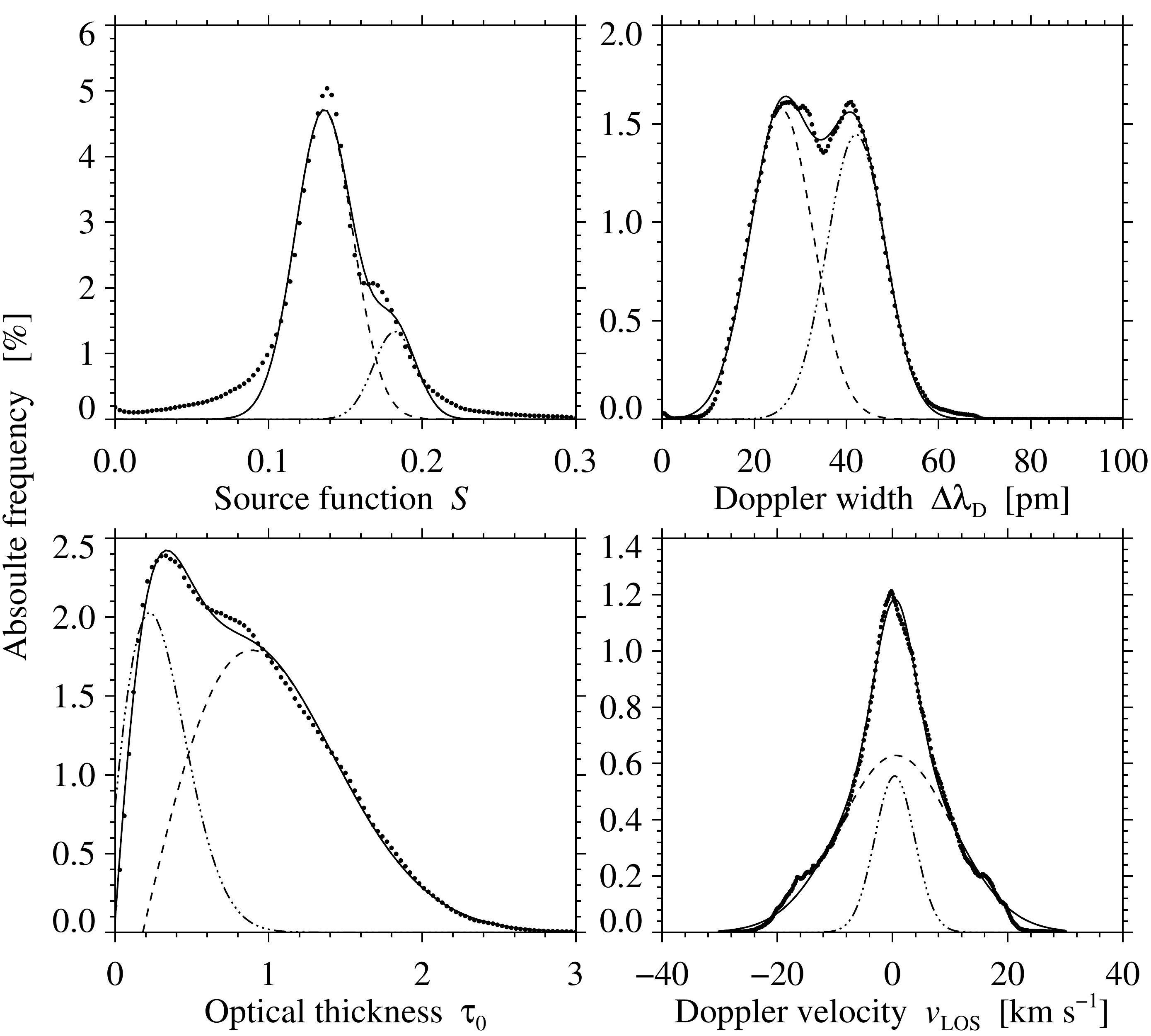}
\caption{Normalized histograms of the CM parameters (dotted)
    depicted in Fig.~\ref{FIG09}. The dashed and dash-dotted curves are double
    Gaussian fits with an exception for $\tau_0$, where two log-normal 
    distributions replace the two Gaussians. The solid curves correspond to the 
    sum of the fitted distributions.}
\label{FIG08}
\end{figure}

Normalize histograms for all CM parameters are given in 
Fig.~\ref{FIG08}.  The distributions for source function $S$ and optical 
thickness $\tau_0$ show a conspicuous `shoulder' on the right side, the one for 
the Doppler width $\Delta\lambda_{D}$ is double-peaked, and the one for the 
velocity $v_\mathrm{LOS}$ possess an extended base. All these characteristics 
are indicative for a sample with two (or more) distinct populations, for example, quiet 
Sun and AFS. In general, the frequency of occurrence is well represented by two 
Gaussians ($S$, $\Delta\lambda_{D}$, and $v_\mathrm{LOS}$). Only in case of the 
optical thickness $\tau_0$, two log-normal distributions are a more appropriate 
model. The mean values $\mu_{1,2}$ and standard deviations $\sigma_{1,2}$ of the 
fit parameters (or the parameter's natural logarithm for $\tau_0$) are given in 
Table~\ref{TAB02} along with the normalized frequency of occurrence $P_1$ of the 
dominant population.

The histograms provide indirect evidence that two features 
with distinct CM parameters are present within the observed FOV. However, they 
do not offer hints, where these features are located, because the distributions 
strongly overlap. Cluster analysis offers a variety of statistical tools to 
identify distinct populations in an $n$-dimensional parameter space. In our case
this means the four-dimensional space of CM parameters (see Figs.~\ref{FIG08} and 
\ref{FIG09}). The number of samples in each histogram is $m = 
200\,188$. The $k$-means clustering algorithm implemented in IDL  
\citep{Everitt2011} requires a priori knowledge about the number of clusters 
$k$. The histograms favor $k=2$. However, the cases $k=3$ and 
$k=4$ are also investigated to validate that two distinct populations are indeed 
sufficient to represent the data. The clustering algorithms assigns the samples 
to a cluster such that the squared distance from the cluster center is minimized 
and the distance between the cluster centers is maximized. The 
clustering algorithm provides additional information to explain the different 
shapes observed in the histogram of Fig.~\ref{FIG08}, meaning 
two peaks for the Doppler width $\Delta\lambda_{D}$, the halo for the LOS 
velocity $v_\mathrm{LOS}$ and the shoulders for the optical thickness$ \tau_0$ 
and the source function $S$.

\begin{figure}[t]
\includegraphics[width=\columnwidth]{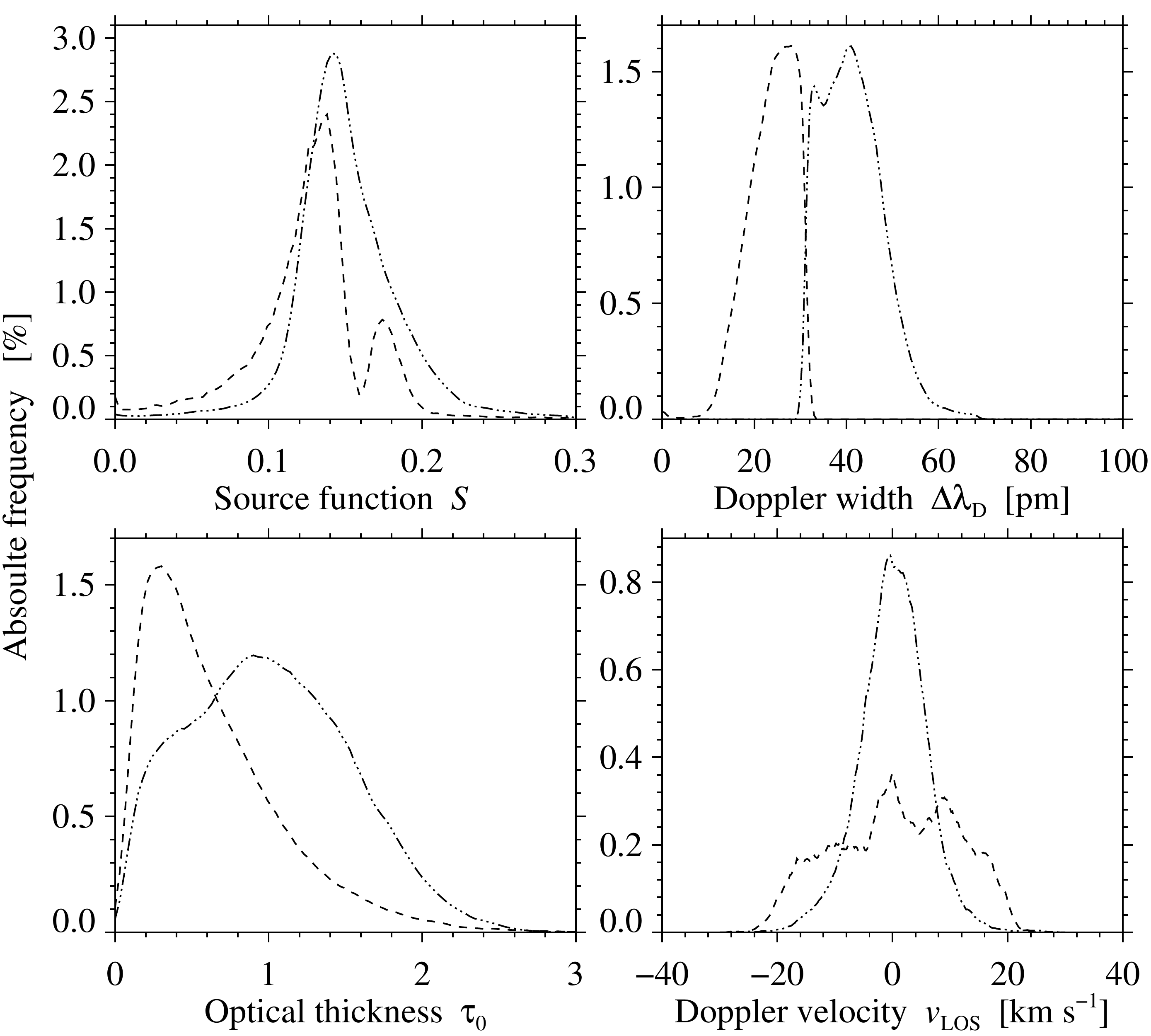}
\caption{Normalized histograms of the CM parameters based on 
    cluster analysis with the same data as in Fig.~\ref{FIG09}. The dashed and 
    dash-dotted curves refer to the two clusters $c_1$ and $c_2$, respectively.}
\label{FIG13}
\end{figure}

Cluster centers or weights $w_i$ are not necessarily the same as the mean values 
$\mu_i$ of the clusters because of the above optimization scheme. In addition to 
the mean values $\mu_i$, we list in Table~\ref{TAB03} the respective standard 
deviations $\sigma_i$ to illustrate the extend and potential overlap of the 
clusters. The size of each population is given by the cluster fraction $c_i$, 
and the normalized histograms for all the CM parameters are 
given in Fig.~\ref{FIG13} for the case $k=2$. 

In the case of just two populations, the Doppler width of the absorption 
profiles $\Delta\lambda_D$ has the strongest discriminatory power. These broad 
absorption profiles also have a higher optical thickness $\tau_0$. On the whole, 
the mean values and standard deviations of the other CM parameters are very 
similar with the exception of the larger standard deviation $\sigma_1$ for the 
velocity $v_\mathrm{LOS}$. This broader distribution is also easily perceived in 
Fig.~\ref{FIG13}. In general, there is a good agreement between the 
distributions in Figs.~\ref{FIG08} and \ref{FIG13}. Only a secondary peak for 
the source function $S$ of cluster $c_1$ and the strong deviation from a 
log-normal distribution of the the optical thickness $\tau_0$ of the cluster 
$c_2$ point to the presence of more than two populations. The cluster fractions 
of $c_1 = 43.3$\% and $c_2 = 56.6$\% are very similar, which however might be 
related to the $k$-means clustering algorithm's tendency to produce clusters of 
similar size.

Increasing the number of clusters to $k=3$ splits the 
population according to the velocity $v_\mathrm{LOS}$ into up- and downflows, 
while at the same time populations~1 and 2 maintain a similar Doppler width 
$\Delta\lambda_D$ as before. Increasing the number of clusters to $k=4$, 
population~3 for the case $k=3$ separates into populations~3 and 4 
according to the Doppler width, which is still significantly larger than for 
populations~1 and 2. In summary, increasing successively the number of clusters 
from $k=2$ to $k=4$, first forks population~1 into up- and downflows and then 
branches population~2 into contrast profiles with narrower and broader Doppler 
width $\Delta\lambda_D$. However, we conclude that two populations are 
sufficient to represent the distributions of the CM parameters, which are most 
easily distinguishable by the Doppler width $\Delta\lambda_D$ and to a lesser 
extend by the optical thickness $\tau_0$ of the cloud material. More than two 
populations likely overinterpret the data because they cannot be associated 
with any particular chromospheric feature.

\begin{figure}[t]
\includegraphics[width=\columnwidth]{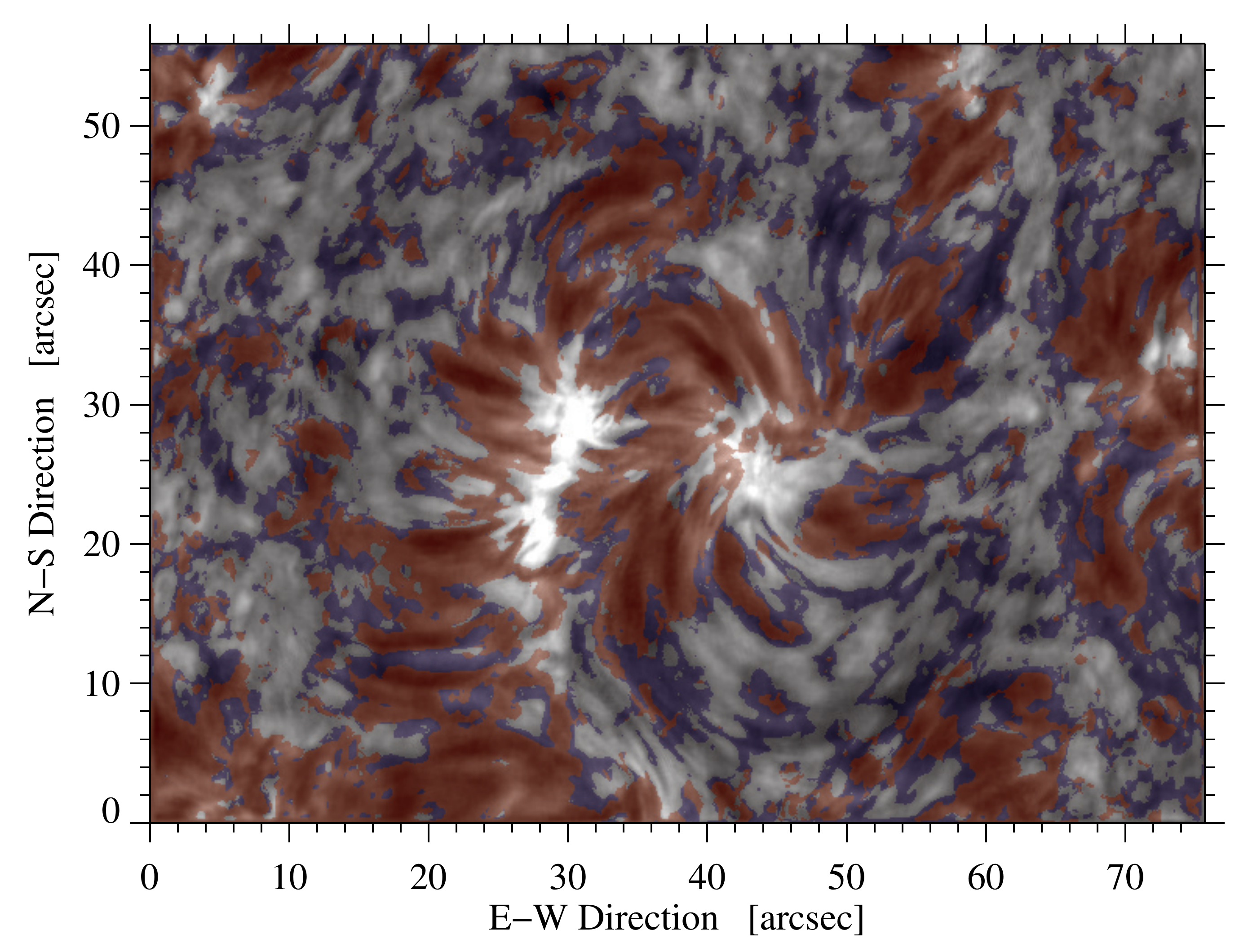}
\caption{Two-dimensional map containing the locations of the
    two clusters $c_1$ (blue) and $c_2$ (red), which were identified in the 
    cluster analysis assuming that the histograms of the CM 
    parameters contain only two distinct populations. 
    The H$\alpha$ line-core intensity image of 
    Fig.~\ref{FIG01} serves as the background. The gray areas indicate the 
    same regions as in Fig.~\ref{FIG09}, where the CM inversions result in 
    mediocre fits.}
\label{FIG14}
\end{figure}

The locations of the two populations are depicted in 
Fig.~\ref{FIG14} in blue and red colors, whereas the gray areas indicate the 
same regions as in Fig.~\ref{FIG09} with mediocre CM fits. The background of the 
figure is the H$\alpha$ line-core intensity image presented in Fig.~\ref{FIG01} 
to assist in matching the two populations to chromospheric features. 
Population~1 (blue) marks the transition to the quiet Sun, where the cloud 
material turns more transparent, that is, $\Delta\lambda_D$ and $\tau_0$ become 
increasingly smaller. Population~2 (red) is thus more representative for the 
small AFS, and the corresponding CM parameters in Tables~\ref{TAB02} and 
\ref{TAB03} should be used in comparisons to values provided in literature. To 
ensure that these two populations are not some artifacts of the CM inversions, 
we examine their $\chi^2$ statistics. The mean values along 
with their standard deviations are $\chi^{2}_{1}=0.064 \pm 0.040$ and 
$\chi^{2}_{2}=0.048 \pm 0.033$, which are essentially the same.

%
%

\section{Discussion}\label{SEC4}

In this study, we analyze a short period in the evolution of the photospheric 
and chromospheric signatures belonging to a small EFR containing two micro-pores 
and a small AFS, where significant changes take place in just a few tens of 
minutes. These micro-pores with sizes of less than one arcsecond, as seen at 
photospheric level, are evolving with time in intensity, shape, and size. Their 
average intensity increased and concurrently the total area decreased, which we 
interpret as a signature of a local decaying process. Fade-out and 
disintegration of strong magnetic elements is also supported by the presence of 
small-scale continuum brigthenings and transient signatures of abnormal 
granulation \citep{DeBoer1992} in the vicinity of the micro-pores. The shape of 
the micro-pores evolves as first reported by \citet{RouppevanderVoort2005} - alternating 
between ribbons and flowers. These authors ascribe the 
dynamics of micro-pores to a fluid-like behavior of magnetic flux, which they 
relate to magnetic features evident in magneto-hydrodynamic simulations 
\citep[e.g.,][]{Carlsson2004, Voegler2005}, see also \citet{Beeck2015} for more 
recent simulation results. At very high spatial and temporal resolution, the 
present study confirms the overall picture of the evolution of magnetic 
concentrations as presented by \citet{Schrijver1997}, among others, in which 
(micro-)pores can split, merge, or vanish. 

The temporal derivatives of the mean total area 
($dA_\mathrm{mp} / dt = -0.22$~km$^2$~s$^{-1}$) and the average intensity 
($dI_\mathrm{mp} / dt = 0.033\,I_0$~s$^{-1}$) can be boundary conditions of 
magnetohydrodynamic simulations on flux emergence in general and particularly in 
solar pore formation \citep[e.g., ][]{Stein2011, Stein2012d, Cameron2007, 
Rempel2014}. Even if the statistical values in the present study only 
represent the evolution of one set of micro-pores, the initial mean intensity 
$0.75\,I_0$ and size  0.8~Mm$^2$ are consistent with other observations and 
simulations. The size dependence of the brightness was reported before in 
observations and simulations \citep[e.g., ][]{Cameron2007, Bonet1995, 
Keppens1996, Mathew2007}. Small pores are not as dark as larger ones. The 
brightness increases when pores decay, and they contract and lose flux 
\citep{Cameron2007}. These authors suggest that  increasing lateral radiative 
heating reduces the pore's size. This effect is compounded by the irregular 
shape of decaying pores.

The LCT results of horizontal flow field around the micro-pores agree 
quantitatively with \citet{Verma2011}. However, the short-averaging time implies 
that the flow speeds more closely reflect motions of individual granules, dark 
micro-pores, and small-scale brightenings. Persistent flows become only apparent 
with longer averaging times \citep{Verma2013}. Yet, the 
significantly reduced flow speed in the micro-pores clearly indicates that the 
magnetic field is sufficiently strong to suppress the convective motions in 
close proximity to the micro-pores and to cancel horizontal motions inside the 
two H$\alpha$ line-core brightenings (see Fig.~\ref{FIG_LCT}). 

The statistical description of micro-pores is still inadequate and limited to 
pores with diameters of about 1~Mm, even when resorting to high-resolution 
Hinode data \citep{Verma2014}. In addition, advanced simulations of emerging 
bipolar pore-like features \citep[e.g.,][]{Stein2011} use larger boxes (about 
50~Mm wide) with larger pores (about 10~Mm diameters) and larger separations of 
the bipoles (about 15~Mm), i.e., they do not necessarily capture the physics and 
dynamics of micro-pores at sub-arcsecond scales. The observed micro-pores are 
`isolated' pores, which represent about 10\% of the pore population 
\citep{Verma2014}. Isolated pores may have their origin in a solar surface 
dynamo \citep{Voegler2007}. Therefore, \mbox{(micro-)}pores in the quiet Sun and 
in active regions are potentially due to different dynamo actions.

The H$\alpha$ line-core intensity, Doppler velocity maps and MDI magnetograms 
reveal a chromospheric arch filamentary structure with two regions of H$\alpha$ 
line-core brightenings and small loops connecting these two areas in a bipolar 
region with opposite polarities. A similar region was also described in 
\citet{Harvey1973}. The width of the dark filaments is a few seconds of arc, and 
their length is around 6\,--\,8\arcsec, which is smaller than the mean lengths 
reported by \citet{Tsiropoula1992}. We find typical aspect ratios of the loops 
of between 1:4 and 1:5. Overall, but with smaller spatial dimensions, the morphology of 
the observed AFS agrees with the findings reported in several studies 
\citep[e.g.,][]{Strous1996, Tziotziou2003, WedemeyerBoehm2009, Zuccarello2009b}.
 
The H$\alpha$ Doppler velocity $v$ at the bright footpoints is mainly downwards 
with a flow speed of up to 12~km~s$^{-1}$. The up- and downflows in the dark 
filamentary system possess an imbalance toward blueshifts with typical mean 
upflow velocities of $\bar{v}_{LT} \approx -1.13$~km~s$^{-1}$ at the loop tops, 
but some upflows reach values close to $-3$~km~s$^{-1}$. Furthermore, the close 
proximity of some up- and downflows taken together with their appearance in 
time-lapse movies support the presence of twisting motions along the loop axis. 
Many other spectroscopic or spectropolarimetry studies have observed AFS in the 
chromospheric absorption line H$\alpha$ and in the Ca\,\textsc{ii}\,H\,\&\,K 
lines \citep{Bruzek1969, Zwaan1985,Chou1988, Lites1998}. The range of the 
downflow Doppler velocities $v$ near the footpoints spans 
30\,--\,50~km~s$^{-1}$, whereas upflows of about 1.5\,--\,22~km~s$^{-1}$ are 
observed at the loop tops. 

In general, the velocities in the present study are much lower than those 
reported by the aforementioned authors. This can be attributed to the smaller 
size of the loop system as compared to previous studies. The higher the 
velocities, the higher the size of the loop systems. The small and isolated 
upflow patches in the present study are closer to the Doppler velocities 
reported in \citet{Lites1998} with observations near the disk center. They 
suggest that the observed AFS is not caused by a monolithic flux rope but the 
result of the dynamical emergence of a filamentary flux bundle. 
The most accepted flux emergence scenario suggest that the 
EFRs are formed by magnetic flux tubes that are transported from the base of the 
solar convection zone (tachocline) to the solar surface by buoyancy 
\citep{Zwaan1987}. Normally, these emerging field appears on the solar surface 
in form of bipolar regions as in our observations known as EFR. The two 
polarities are commonly the footpoints of an $\Omega$-loops system rising to the solar 
surface even up to the solar corona \citep[e.g., ][]{Schmieder2004, Strous1999, 
Strous1996} Furthermore, at chromospheric level, flows occur within the AFS 
accompanied by draining material from the H$\alpha$ loops toward the 
footpoints, where the higher downflows reside. In the case of \citet{Lites1998}, 
this occurs where pores and sunspots form. In our case, only one footpoint shows 
an intensity signature - micro-pores evolving in the photosphere. 
Another explanation of the physical processes in loops 
connecting the two footpoints is an upflow that starts in the middle of 
two regions with opposite magnetic flux, in our case, in the middle and upper 
part of the two footpoints (magnetic configuration of the case~C in Fig.~12 of 
\citet{Lee2000}. The origin of the upflow events is then explained by  magnetic 
reconnection. The mass moves upward through the small loops and falls down in 
other areas far away.

We have presented maps of four CM parameters and the corresponding normalized 
histograms based on CM inversions of H$\alpha$ contrast 
profiles (Figs.~\ref{FIG08} and \ref{FIG09}), i.e., source function $S$, Doppler 
width $\Delta\lambda_D$, optical thickness $\tau_{0}$, and velocity of the cloud 
material $v_\mathrm{LOS}$. The inversion scheme provided well matched profiles 
for around 60\% of the entire FOV. The population of the inverted AFS contrast 
profiles covered only about 2\% of the entire FOV. Failed CM inversions are 
mainly related to much of the quiet-Sun regions and the H$\alpha$ line-core 
brightenings, where the underlying assumptions of the model are violated.
 
The inversion results for quiet-Sun regions, dark filamentary features, and the 
ensemble average show in general similar values. However, the Doppler width and 
the optical thickness of the dark chromospheric filaments 
are much larger than in the quiet-Sun regions, and the velocity $v_\mathrm{LOS}$ 
possesses, similar to the quiet-Sun regions, a small upflow on average. 
We note that the average quiet-Sun Doppler velocity is 
based on the center-of-gravity methods not the CM inversions. Furthermore, the 
velocity $v_\mathrm{LOS}$ of the entire FOV is on average redshifted because of the 
downflows being present in the EFR and the AFS, with the exception of the loop 
tops. In the CM maps it is clearly seen that the parameters $S$, 
$\Delta\lambda_{D}$, and $\tau_0$ reach the largest values in proximity to the 
H$\alpha$ line-core brightenings. In regions lacking any filamentary structure, 
the values are much lower. The results of the CM parameters in 
Table~\ref{TAB02} are very similar to the ones presented by \citet{Bostanci2011} 
for dark mottles. They are also close to the results of AFSs investigated 
by \citet{Alissandrakis1990}, whereas the source function $S$ and optical 
thickness $\tau_{0}$  deviate from the results presented by \citet{Bostanci2010, 
Lee2000}. 

The AFS in the present study certainly belongs to the smallest observed 
chromospheric filaments. Only mini-filaments \citep{Wang2000b} are smaller. 
Interestingly, the CM inversion results are similar as compared to a giant 
filament investigated by \citet{Kuckein2016}. As expected, their mean optical 
thickness $\tau_{0}$ is larger than our average inferred values.

The normalized histograms suggest that the CM parameters represent two different 
populations, which is corroborated by the cluster analysis and the computed absolute contrast that 
clearly distinguishes between features with higher contrast belonging to the AFS and those being part of 
quiet-Sun regions or the interface between the quiet Sun and the AFS. We did expect 
that for similar observations with a small FOV containing AFSs two populations may appear 
again all over the solar disk and distinguish between dark AFS and the regions of the quiet-Sun and the 
transition to the quiet-Sun. We note that in several studies the authors select only region of interest, 
that is, the features with high contrast and they do not invert the quiet-Sun regions, \citep[e.g., 
][]{Kuckein2016, Contarino2009}. For larger FOVs that may contain different features like active regions, 
filaments, quiet-Sun, rosette structures, etc., we foresee that CM inversions together with 
cluster analysis will separate these chromospheric features in different populations 
based on the four CM parameters.  

%
%

\section{Conclusions}\label{SEC6}

Photospheric and chromospheric structures like sunspots and 
pores cover a broad range of spatial scales. Sunspots come into existence as 
pores, grow, and potentially assemble into large (complex) active regions. In 
the chromosphere, cool plasma is suspended by the magnetic field in structures 
ranging in size from mini-filaments (a few megameters) to giant and polar crown 
filaments (several hundred megameters). This raises the questions if scaling 
laws exist characterizing the physical properties of these phenomena. In EFRs 
with AFSs both photospheric and chromospheric properties are relevant. Exploring 
the lower end of their spatial scales requires large-aperture solar telescopes 
with dedicated instruments. As a consequence the observed FOV is small, 
time series cover at most a few hours, and synoptic observations building up a 
database are often impracticable. Therefore, case studies are the only means of 
advancing our knowledge of EFRs and AFSs at sub-arcsecond scales.

In the present study, we used the G\"ottingen Fabry-P\'erot 
Interferometer to investigate a very small EFR with its accompanying AFS. It is almost 
inconceivable that systems with a clear bipolar magnetic structure exist that are 
smaller than this. Mini-filaments are not considered in this context 
because they reside at the polarity inversion line (PIL) between opposite 
magnetic polarities \citep{Wang2000b}, thus representing a different magnetic 
field topology. In our observations, the AFS connects the micro-pores to a 
quiet-Sun region of opposite polarity flux, that is, the small-scale filaments are 
perpendicular to the PIL. The filaments or dark fibrils are a clear indication 
of a newly emerging flux in the form of $\Omega$-loops. The magnetograms 
represented in Fig.~\ref{BIPOLE} show that the trailing polarity is formed earlier than 
the leading polarity and that it is formed earlier than the leading polarity. However, the leading 
polarity is more compact and survives longer. This asymmetry is a typical property 
of EFRs, also for larger ones  \citep{vanDrielGesztelyi1990}. The decay of the EFR is characterized by flux 
cancellation and fragmentation. The flux system emerged and then submerged as a 
whole similar to a larger EFR studied in \citet{Verma2016}. Flux fragmentation 
indicate that several $\Omega$-loops and not just one monolithic loop connects 
the opposite polarities, which is also reflected in the multiple dark fibrils 
constituting the AFS.

The statistical description of micro-pores and flux emergence 
at spatial scale below 1~Mm remains challenging. We presented, among others, 
evolution timescales for the area of micro-pores, their mean intensity, and the 
separation of footpoints, which match previous observations and simulations, 
when taking into acount the size of the observed EFR and AFS. In the 
chromosphere, upflows are mainly observed at the loop tops and from there the 
cool plasma drains toward the footpoints. Generally, the LOS velocities 
inferred in the loops are two to four times lower than for larger AFSs. Buoyancy 
and curvature of the rising or submerging $\Omega$-loops is here the determining 
factor but a statistically more meaningful sample is clearly needed to derive 
scaling laws. Pores occur on spatial scales ranging from sub-arcsecond to about 
ten arcseconds. Thus, comparing observations of EFRs and AFSs in this spatial 
domain with simulations of flux emergence and decay will be very beneficial for 
our understanding of the dynamic interaction of magnetic fields with the 
surrounding plasma.

A shortcoming of the present study was the lack of 
high-resolution spectropolarimetric observations, which motivates follow-up 
observations with the GREGOR Fabry-P\'erot Interferometer 
\citep[GFPI,][]{Puschmann2012} and the GREGOR Infrared Polarimeter 
\citep[GRIS,][]{Collados2012} at the 1.5-meter GREGOR solar telescope 
\citep{Schmidt2012}. Both instruments are capable of full-Stokes polarimetry and 
allow multi-wavelength observations of multiple spectral lines (e.g., H$\alpha$ 
with the GFPI and the spectral region around the He\,\textsc{i} triplet at 
1083~nm with GRIS). In addition to magnetic field information, the height 
dependence of physical properties in EFRs and AFSs becomes thus accessible.

%
%

\begin{acknowledgements}
The Vacuum Tower Telescope (VTT) at the Spanish Observatorio del Teide of the 
Instituto de Astrof{\'\i}sica de Canarias (IAC) is operated by the 
Kiepenheuer-Institut f\"ur Sonnenphysik (KIS) in Freiburg. The 
Solar and Heliospheric Observatory (SoHO) is a project of international 
cooperation between the European Space Agency (ESA) and the National Aeronautics 
and Space Administration (NASA). SJGM is grateful for financial support from 
the Leibniz Graduate School for Quantitative Spectroscopy in Astrophysics, a 
joint project of the Leibniz Institute for Astrophysics Potsdam (AIP) and the 
Institute of Physics and Astronomy of the University of Potsdam (UP). 
N.B.G. acknowledges financial support by the Senatsausschuss of 
the Leibniz-Gemeinschaft, Ref.-No. SAW-2012-KIS-5. CD acknowledges support by 
grant DE 787/3-1 of the German Science Foundation (DFG). We thank Drs.\ M.~L\"ofdahl 
and T.~Hillberg for their support in installing the latest version of MOMFBD. We are 
indebted to Dr.\ M.~Verma, A.~Diercke, Dr.\ C.~Kuckein, and Dr.\ H.~Balthasar for 
agreeing to critically read the manuscript and offering comments and suggestions. 
We are grateful to Dr.\ A.\ Hofmann for his assistance during the observations.
\end{acknowledgements}


\bibliographystyle{aa}
\bibliography{aa-jour,cdenker}

\end{document}